\documentstyle[preprint,prc,aps,tighten,floats,epsf]{revtex}

\begin{document}
%
%
\def\symdef#1#2{\gdef#1{#2}}   
\symdef{\alphabar}{\overline\alpha}
\symdef{\alphabarp}{\overline\alpha\,{}'}
\symdef{\Azero}{A_0}

\symdef{\bzero}{b_0}

\symdef{\Cs}{C_{\rm s}}
\symdef{\cs}{c_{\rm s}}
\symdef{\Cv}{C_{\rm v}}
\symdef{\cv}{c_{\rm v}}

\symdef{\d}{{\rm d}}
\symdef{\Deltaevac}{\Delta{\cal E}_{\rm vac}}

\symdef{\ed}{{\cal E}}
\symdef{\edk}{{\cal E}_k}
\symdef{\edkzero}{{\cal E}_{k0}}
\symdef{\edv}{{\cal E}_{\rm v}}
\symdef{\edvphi}{{\cal E}_{{\rm v}\Phi}}
\symdef{\edvphizero}{{\cal E}_{{\rm v}\Phi 0}}
\symdef{\edzero}{{\cal E}_{0}}
\symdef{\Efermistar}{E_{{\scriptscriptstyle \rm F}}^\ast}
\symdef{\Efermistarzero}{E_{{\scriptscriptstyle \rm F}0}^\ast}
\symdef{\etabar}{\overline\eta}
\symdef{\ezero}{e_0}

\symdef{\fpi}{f_\pi}

\symdef{\gA}{g_A}
\symdef{\gpi}{g_\pi}
\symdef{\grho}{g_\rho}
\symdef{\gs}{g_{\rm s}}
\symdef{\gv}{g_{\rm v}}

\symdef{\fm}{\mbox{\,fm}}

\symdef{\infm}{\mbox{\,fm$^{-1}$}}

\symdef{\kappabar}{\overline\kappa}
\symdef{\kfermi}{k_{{\scriptscriptstyle \rm F}}}
\symdef{\kfermizero}{k_{{\scriptscriptstyle \rm F}0}}
\symdef{\Kzero}{K_0}

\symdef{\lambdabar}{\overline\lambda}
\symdef{\lzero}{l_{0}}

\symdef{\MeV}{\mbox{\,MeV}}
\symdef{\mpi}{m_\pi}
\symdef{\mrho}{m_\rho}
\symdef{\ms}{m_{\rm s}}
\symdef{\Mstar}{M^\ast}
\symdef{\Mstarzero}{M^\ast_0}
\symdef{\mv}{m_{\rm v}}
\symdef{\mzero}{{\rm v}_{0}}

\symdef{\Phizero}{\Phi_0}
\symdef{\psibar}{\overline\psi}
\symdef{\pvec}{{\bf p}}

\symdef{\rhoB}{\rho_{{\scriptscriptstyle \rm B}}}
\symdef{\rhoBzero}{\rho_{{\scriptscriptstyle \rm B}0}}
\symdef{\rhos}{\rho_{{\scriptstyle \rm s}}}
\symdef{\rhospzero}{\rho'_{{\scriptstyle {\rm s} 0}}}
\symdef{\rhoszero}{\rho_{{\scriptstyle {\rm s}0}}}
\symdef{\rhozero}{\rho_0}

\symdef{\Szero}{S_0}

\symdef{\umu}{u^\mu}
\symdef{\Ualpha}{U_{\nu}}
\symdef{\Uzero}{U_0}
\symdef{\Uzerop}{U_0'}
\symdef{\Uzeropp}{U_0''}

\symdef{\vecalpha}{{\bbox{\alpha}}}
\symdef{\veccdot}{{\bbox{\cdot}}}
\symdef{\vecnabla}{{\bbox{\nabla}}}
\symdef{\vecpi}{{\bbox{\pi}}}
\symdef{\vectau}{{\bbox{\tau}}}
\symdef{\vecx}{{\bf x}}
\symdef{\Vzero}{V_0}

\symdef{\wzero}{w_0}
\symdef{\Wzero}{W_0}

\symdef{\zetabar}{\overline\zeta}
%
%
%
%
%
\def\capcrunch{%
    \setlength{\baselineskip}{0.2\baselineskip}
    \setlength{\hangindent}{0.04\textwidth}
    \setlength{\hsize}{0.84\textwidth}
    }

%


\preprint{\vbox{\hfill IU/NTC\ \ 95--16\\
                \null\hfill OSU--95-0326}}

\title{Analysis of Chiral Mean-Field Models for Nuclei}

\author{R. J. Furnstahl}
\address{Department of Physics \\
         The Ohio State University,\ \ Columbus, Ohio\ \ 43210}
\author{Brian D. Serot}
\address{Department of Physics and Nuclear Theory Center \\
         Indiana University,\ \ Bloomington, Indiana\ \ 47405}
\author{Hua-Bin Tang%
  \footnote{Present address: School of Physics and Astronomy,
     University of Minnesota, Minneapolis, MN\ \ 55455.}}
\address{Department of Physics \\
         The Ohio State University,\ \ Columbus, Ohio\ \ 43210}
%

%
\date{November, 1995}
\maketitle
\begin{abstract}

An analysis of nuclear properties based on a relativistic energy
functional containing Dirac nucleons and classical scalar and
vector meson fields is discussed.
Density functional theory implies that
this energy functional can include many-body effects that go
beyond the simple Hartree approximation.
Using basic ideas from effective field theory, a systematic
truncation scheme is developed for the energy functional, which
is based on an expansion in powers of the meson fields and
their gradients.
The utility of this approach relies on the observation that the
large scalar and vector fields in nuclei are small enough
compared to the nucleon mass to provide useful expansion parameters,
yet large enough that exchange and correlation corrections to
the fields can be treated as minor perturbations.
Field equations for nuclei and nuclear matter are obtained by
extremizing the energy functional with respect to the field variables,
and inversion of these field equations allows one to express the
unknown coefficients in the energy functional directly in terms of
nuclear matter properties near equilibrium.
This allows for a systematic and complete study of the parameter space,
so that parameter sets that accurately reproduce nuclear observables
can be found, and models that fail to reproduce nuclear properties can
be excluded.

Chiral models are analyzed by considering specific lagrangians that
realize the spontaneously broken chiral symmetry of QCD in different
ways and by studying them at the Hartree level.
The resulting energy functionals are special cases of the general
functional considered earlier.
Models that include a light scalar meson playing a dual role as the
chiral partner of the pion and the mediator of the intermediate-range
nucleon--nucleon interaction, and which include a ``Mexican-hat'' potential,
fail to reproduce basic ground-state properties of nuclei at the
Hartree level.
In contrast, chiral models with a {\em nonlinear\/} realization of the
symmetry are shown to contain the full flexibility inherent in the general
energy functional and can therefore successfully describe nuclei.
\end{abstract}


\narrowtext


\section{Introduction}
\label{sec:Intro}

There is a long history of attempts to unite  relativistic
mean-field phenomenology with manifest chiral symmetry.
In particular, it has been tempting to build upon the linear
sigma model \cite{SCHWINGER57,GELLMANN60,LEE72},
where a light scalar meson plays a dual role as the chiral
partner of the pion {\it and\/} the mediator of the intermediate-range
nucleon--nucleon (NN) attraction
\cite{LEE74,KERMAN74,NYMAN76a,NYMAN76b,MATSUI82,BOGUTA83,JACKSON83,%
SARKAR85,KUNZ86,KUNZ87,BOGUTA89,FOMENKO92,FOMENKO95}.
In earlier work \cite{FURNSTAHL93}, we surveyed a broad class of these
chiral hadronic models and observed that they fail to reproduce
basic properties of finite nuclei at the Hartree level.
In this paper, we demonstrate the generic failure of this type of model
using a more complete approach \cite{BODMER91}
for analyzing the relationship between model parameters and nuclear
observables.
We also illustrate the characteristics of chiral models that {\it can\/}
successfully describe finite nuclei.

We base our analysis on an energy functional,
which depends on valence-nucleon Dirac wave functions
and classical scalar and vector meson fields.
Extremizing the functional leads to coupled equations
for finite nuclei and nuclear matter
[see Eqs.~(\ref{eq:efunctional})--(\ref{eq:diraceq})].
The successes of
relativistic mean-field models have shown that these variables
(nucleon, scalar, vector) allow an efficient and natural description
of bulk and single-particle nuclear properties
\cite{HOROWITZ81,SEROT86,REINHARD86,FURNSTAHL87,REINHARD89,GAMBHIR90,%
SEROT92,SHARMA93a,SHARMA93b}.

Although the energy functional contains classical meson fields,
this framework can accommodate physics beyond the simple Hartree
(or one-baryon-loop) approximation.
This is achieved by combining aspects of both
density functional theory
(DFT) \cite{DREIZLER90,SPEICHER92,SCHMID95,ENGEL95}
and effective field theory (EFT) \cite{WEINBERG67,GEORGI93,WEINBERG95}.
In a DFT formulation of the relativistic nuclear
many-body problem, the central object is an energy functional of scalar and
vector densities (or more generally, vector four-currents).
Extremization of the functional gives rise to Dirac equations
for occupied orbitals
with {\it local\/} scalar and vector potentials,
not only in the Hartree approximation, but in the general case as well.%
\footnote{Note that the Dirac eigenvalues do not correspond precisely
to physical energy levels in the general case \protect\cite{DREIZLER90}.}
Rather than work solely with the densities,
we can introduce auxiliary variables corresponding to the local potentials,
so that the functional depends also on meson fields.
The resulting DFT formulation takes the form of a Hartree calculation, but
correlation effects can be included,
{\it if\/} the proper density functional can be found.
Our procedure is analogous to the well-known Kohn--Sham \cite{KOHN65}
approach in DFT, with the local meson fields playing the role of Kohn--Sham
potentials; by introducing nonlinear couplings between these fields, we can
implicitly include density dependence in the single-particle potentials.

Moreover, by introducing the meson fields, we can incorporate the ideas
of EFT.
The exact energy functional has kinetic energy and Hartree parts
(which are combined in the relativistic formulation) plus
an ``exchange-correlation'' functional, which is a
nonlocal, nonanalytic functional of the densities that contains all the
other many-body and relativistic effects.
We do not try to construct the
latter functional explicitly from a lagrangian (which
would be equivalent to solving the full many-body problem), nor do we
attempt here to construct an explicit functional using standard many-body
techniques \cite{SCHMID95}.
Rather, we approximate the functional using an expansion in classical meson
fields and their derivatives.
The parameters introduced in the expansion can be fit to experiment, and
if we have a systematic way to truncate the expansion, the framework is
predictive.
Thus a conventional mean-field energy functional fit directly to nuclear
properties, if allowed to be
sufficiently general, will automatically incorporate
effects beyond the Hartree approximation, such as those due to
short-range correlations.

We rely on the special characteristics of nuclear ground states
in a relativistic formulation, namely, that the mean
scalar and vector potentials $\Phi$ and $W$ are large on nuclear energy scales
but are small compared to the nucleon mass
$M$ and vary slowly in finite nuclei\cite{BODMER91,FURNSTAHL95}.
This implies that the ratios $\Phi/M$ and $W/M$ and the
gradients $|\nabla\Phi|/M^2$ and $|\nabla W|/M^2$
are useful expansion parameters.
Moreover, as is illustrated in Dirac--Brueckner--Hartree--Fock (DBHF)
calculations \cite{HOROWITZ87,TERHAAR87,MACHLEIDT89},
the scalar and vector potentials (or self-energies) are nearly
state independent and are dominated by the Hartree contributions.
Thus the Hartree contributions to the energy functional should dominate,
and an expansion of the exchange-correlation functional in terms of mean
fields should be a reasonable approximation.
This ``Hartree dominance'' also implies that it should be a good
approximation to associate the single-particle Dirac eigenvalues with the
observed nuclear energy levels, at least for states near the Fermi
surface \cite{DREIZLER90}.
Of course, the mean-field expansion cannot accommodate all of the nonlocal and
nonanalytic aspects of the exchange-correlation functional;
however, we reserve a complete discussion of the limits of this approach
for future work \cite{FURNSTAHL96}.

The mean-field energy functional is specified by the values of various
coupling constants and masses.
These parameters can be constrained in two ways.
First,
the energy functional provides  a framework
for relating model parameters to appropriate
nuclear observables.
This translates empirical nuclear properties into conditions on acceptable
parametrizations.
In practice, nuclei provide rather stringent constraints.
On the other hand,
the realization of QCD symmetries (such as chiral symmetry)
in a candidate model lagrangian
may impose relationships between parameters in the corresponding
Hartree approximation to the energy functional.
Our goal in this work is to test whether the two sets of constraints are
compatible.

We begin by analyzing a general energy functional to determine
the characteristics that generate successful phenomenology.
An essential element is a plausible truncation scheme, so that the
functional does not entail an unrestricted number of parameters.
As indicated earlier,
our basic expansion parameters are $\Phi/M$ and $W/M$, where $\Phi$
and $W$ correspond to the potentials in the single-particle Dirac equations
[see Eq.~(\ref{eq:diraceq})].
We define success through a set of nuclear observables that should be
quantitatively reproduced in any useful description of ground-state
properties.
These observables include: 1) nuclear shape properties, such as
charge radii and charge densities, 2) nuclear
binding-energy systematics,
and 3) single-particle properties such as level spacings and orderings,
which reflect spin-orbit splittings and shell structure.
If one recalls that the Kohn--Sham approach is formulated to reproduce
precisely the ground-state density, and that the Hartree contributions
are expected to dominate the Dirac single-particle potentials,
{\it these observables are precisely the ones for which meaningful comparisons
with experiment should be possible.\/}
Moreover, experience has shown that these observables can be replaced by a
set of nuclear matter  properties plus constraints on the meson
masses \cite{HOROWITZ81,SEROT86,REINHARD86,REINHARD89,GAMBHIR90}.
While the parameters of a general functional are not completely determined
by requiring these properties to be reproduced, regions in parameter space
that generate realistic nuclei can clearly be identified.

The second part of our analysis
is to examine specific chiral models that adopt a particular
realization of chiral symmetry at the lagrangian level {\it and\/} a
particular mechanism for spontaneous symmetry breaking.
These two features restrict the accessible regions of parameter space, and
we can simply compare with the regions determined in the first part of our
analysis to see if there is an overlap.
We stress that in this part of the analysis, we actually start with an
explicit lagrangian and construct the energy functional by working strictly
in the Hartree approximation; thus, the effects of correlations in
these chiral models, for example, are beyond our analysis.
Nevertheless, the Hartree energy functionals are simply special cases of
the general functional considered above, so we can apply our earlier results.

The most important ingredient is a procedure that can systematically connect
the nuclear observables and the model parameters.
In previous analyses, the mean-field parameters have been determined
either by fitting nuclear matter equilibrium properties plus some input from
finite nuclei ({\em e.g.}, the rms radius of $^{40}$Ca) or
by optimizing the predictions of selected
experimental observables across the Periodic Table (using a $\chi^2$
measure, for example).
These procedures
are useful if the goal is to find a good description of nuclei for a
given model, but not if we want to rule out a class of models.
In particular, it is difficult to search through the parameter space to
determine where the boundaries of acceptable parametrizations lie.
This severely limits qualitative insight and solid conclusions.

The first approach was followed in our previous studies of linear sigma
models: we explored the chiral-model parameter space by
fitting to nuclear matter equilibrium properties
and then by calculating finite nuclei for what we hoped was an
exhaustive range of parameter sets.
In the present work, we invert this analysis by adapting the approach
developed by Bodmer \cite{BODMER91},
which enables us to use basic nuclear properties, derived from the
observables, {\it to solve directly for the model parameters}.
Since there are only a few basic properties that provide meaningful
constraints, the resulting conditions are undercomplete for the most general
functionals, and there are
families of acceptable solutions defined by regions in the parameter space.
Specific models, however, introduce relations among some parameters and
set others to zero, so that in the accessible region of parameter
space there may be no acceptable descriptions of nuclei.
Thus one way a model can fail is if there is no overlap between the
accessible and desirable regions of parameter space.

There is another way a model can fail, which arises from the nonlinear
meson interactions in the energy functional and in the resulting meson
field equations.
For some models and some parameter sets, one can find a nuclear ground state
that has the appropriate properties at equilibrium, but this ground state
exists only over a narrow range of nuclear densities.
If the ground state fails to exist at densities that have clearly been
obtained experimentally (say, slightly higher than normal nuclear matter
density), the disappearance of the model ground state must also be
considered a failure of the model.
We will make these considerations more precise as we proceed.

The results of the present analysis
solidify our previous conclusion \cite{FURNSTAHL93}
that chiral hadronic models built upon
the conventional linear sigma model cannot reproduce observed properties
of finite nuclei in the Hartree approximation.
The main problems in these linear sigma models arise from
the nonlinear terms in the ``Mexican-hat'' potential, which serves to
precipitate spontaneous symmetry breaking, and from the large pion coupling.
The consequences are a scalar meson mass that is too large and
density-dependent forces (in the form of scalar self-couplings)
with the wrong systematics.
These problems are not remedied by including one-baryon-loop vacuum
corrections or by adding parameters that generalize the model.

On the other hand, chiral models that {\it are\/} phenomenologically
successful can be constructed.
As an example, we discuss
a model that was introduced in a recent paper \cite{FURNSTAHL95},
which features a nonlinear realization of chiral symmetry.

For a model to be viable, it should also exhibit ``naturalness'' in the
fitted parameters.
Here, naturalness implies that the coefficients of the various terms in the
energy functional, when expressed in appropriate dimensionless form,
should all be of order unity.
This means that we can anticipate the approximate magnitude of
contributions to the energy functional (at least up to moderate nuclear
densities) and thereby motivate a suitable expansion and truncation scheme
for the functional; if the coefficients are natural, the omitted terms
will be numerically unimportant.
The naturalness of our effective field theory also implies that one should
include all possible terms (that is, those allowed by the symmetries) through
a given order of truncation; it is {\it unnatural\/} for some coefficients
to vanish without an appropriate symmetry argument.
One consequence is that nonlinear interactions of the vector
field should play an important role in providing a good and natural fit to
nuclei, and we indeed find that this is the case.
We stress this point because nearly all investigations of relativistic
mean-field phenomenology to
date \cite{BOGUTA77,REINHARD86,FURNSTAHL87,REINHARD89,GAMBHIR90,%
SHARMA93a,SHARMA93b} have arbitrarily restricted consideration
to scalar self-interactions only.\footnote{%
Some previous analyses were based on {\em renormalizable\/} scalar--vector
theories, for which such restrictions are appropriate.
If renormalizability is abandoned, however, as in an effective field theory,
there is no reason to omit vector--vector and scalar--vector interactions.
(See the
comments below on vector self-interactions, spin content, and causality.)}
The quartic vector self-coupling has been included in phenomenological
calculations only recently \cite{BODMER91,GMUCA92,KOUNO95}.

There are many studies in the literature that have characterized
the nature of phenomenologically successful mean-field models, including
Refs.~\cite{REINHARD86,FURNSTAHL87,REINHARD89,GAMBHIR90,%
SHARMA93a,SHARMA93b}.
We echo many of the common conclusions, such as
the necessity for a small nucleon effective mass
($0.58 \lesssim \Mstar/M \lesssim 0.64$) at equilibrium density and for a
scalar meson mass of roughly $500\,{\rm MeV}$.
Nevertheless, since previous analyses typically determined the model
parameters through a fitting procedure that simply calculates nuclei
repeatedly until a ``good fit'' is obtained, connections between nuclear
observables and the resulting parameters are somewhat obscure.
Through our analysis, we try to demystify the connections between the
observables and the parameters, so that the structure of the resulting
energy functionals can be better understood.

The original motivation for this work was to expand upon our previous
analyses of linear chiral models \cite{FURNSTAHL93} and thereby solidify
our earlier conclusions.
In carrying out this program, however, we were forced to extend our
mean-field machinery in several different directions.
It is important to emphasize these new aspects, which are more general than
the chiral-model analysis, and which in fact comprise the most important
results in this paper.
First, by starting from an energy functional containing baryons and
classical scalar and vector fields, we have a framework that goes beyond
the simple Hartree approximation.
This follows because we can interpret the analysis in the context of density
functional theory and an associated (relativistic) Kohn--Sham approach.
Second, although it is possible to vary the parameters in the energy
functional until desirable properties of nuclear matter and finite nuclei
are obtained, it is more efficient to invert the field equations and
express the model parameters directly in terms of the desired observables.
This allows for a systematic investigation of the parameter space and
clarifies the relationships between observables and model parameters.
Finally, by applying ideas from effective field theory, such as the
importance of naturalness and a suitable expansion scheme, to the
construction of the ground-state energy functional, we find that it is
important and necessary to include all allowable terms in the functional
at the chosen level of truncation.
This modern view of relativistic quantum field theory \cite{WEINBERG95}
generalizes earlier approaches based on renormalizable
theories \cite{WALECKA74,CHIN77,SEROT86,SEROT92}
and allows for a more unified
discussion of relativistic approaches to the nuclear many-body problem.

The paper is organized as follows:
In Section~\ref{sec:setup}, we consider a  general energy functional
(not manifestly chiral) and
characterize phenomenologically successful models.
In Section~\ref{sec:Bodmer}, we show how
empirical properties of nuclear matter and finite nuclei can be used
directly to determine or to constrain the
parametrizations of candidate energy functionals.
We also illustrate how to map out the parameter space.
In Section~\ref{sec:sigma}, we
specialize the analysis to models built on the linear sigma model
with a ``Mexican-hat'' potential and demonstrate their generic failure at the
Hartree level.
In Section~\ref{sec:tang},
we consider a chiral model that successfully describes finite nuclei.
Section~\ref{sec:discuss} has some additional
discussion, including
the general strategy that justifies the expansion and truncation scheme.
Section~\ref{sec:summary} is a summary.


\section{Energy Functionals and Nuclear Properties}
\label{sec:setup}

Our first goal is to construct an energy functional whose extremization
accurately reproduces bulk and single-particle nuclear observables.
Experience with relativistic mean-field models shows that scalar and vector
mesons with appropriate masses lead to NN interactions with the desired
ranges, and that by introducing nonlinearities in the meson fields, we can
include density dependence in these
interactions \cite{BOGUTA77,HOROWITZ81,BODMER91}.
We assume that the nuclear ground states have good parity and are invariant
under time reversal,
 which implies that a classical pion field does not
appear\footnote{When odd--$A$ nuclei are studied using
mean-field models, time-reversal invariance is broken explicitly, which leads
to pion mean fields.}~\cite{FURNSTAHL87}.
Furthermore, we retain only the valence nucleons explicitly.
This does not mean that we are {\it neglecting\/} pionic or vacuum effects,
but rather that they are implicitly contained in
 the coefficients of the energy
functional.
This is discussed for one-loop vacuum contributions
in Ref.~\cite{FURNSTAHL95} and in Sec.~\ref{sec:tang};
a more general discussion will be given elsewhere \cite{FURNSTAHL96}.

As emphasized in the Introduction, while our energy functional
can be interpreted as arising from a hadronic lagrangian treated
in the Hartree approximation, it can
also be interpreted as an  approximation to
a general density functional that incorporates all correlation effects.
This approach is particularly compelling because of the dominance of
Hartree effects in the relativistic approach.
This claim is supported by Dirac--Brueckner--Hartree--Fock (DBHF) calculations,
which indicate that exchange terms and short-range correlations do not
significantly change the size of the Hartree self-energies
nor introduce strong state dependence (at least for occupied
states) \cite{HOROWITZ84,HOROWITZ87}.
By determining the model parameters from finite-density bulk and
single-particle observables instead of from NN scattering,
and by explicitly allowing
meson nonlinearities that generate additional density dependence,
we automatically include the most important effects of correlations.
In fact, one can choose the nonlinear meson parameters so that the mean
fields reproduce the scalar and vector self-energies obtained in a
DBHF calculation \cite{GMUCA92}.

We start with a mean-field energy functional for spherically or axially
symmetric ground states that generalizes the functional used in
Ref.~\cite{FURNSTAHL87}.
It depends on a set of single-particle Dirac
wave functions $\Ualpha(\vecx)$
(labeled by quantum numbers $\nu$) for the occupied valence orbitals
and on classical, static meson fields $\phi(\vecx)$ and $\Vzero(\vecx)$.
It can be written as
\begin{eqnarray}
  E[\{\Ualpha\},\phi,\Vzero] &=& \int\! {\rm d}^3x \, \biggl(\,
  \sum_{\nu}^{\rm occ} \Ualpha^\dagger(\vecx) \bigl\{
  -i\vecalpha\veccdot\vecnabla + \beta[M - \gs\phi(\vecx)]
   + \gv\Vzero(\vecx) \bigr\}  \Ualpha(\vecx)
   \nonumber \\[4pt]
  & & \quad\null
+  \frac12 \bigl\{ [\vecnabla\phi(\vecx)]^2 + \ms^2[\phi(\vecx)]^2] \bigr\}
 - \frac12 \bigl\{ [\vecnabla\Vzero(\vecx)]^2 + \mv^2[\Vzero(\vecx)]^2
            \bigr\}
  \nonumber \\[4pt]
  & & \quad\null
	    + \frac{1}{3!}\kappa [\phi(\vecx)]^3
            + \frac{1}{4!}\lambda [\phi(\vecx)]^4
  - \frac{1}{4!}\zeta  [\gv\Vzero(\vecx)]^4
   \nonumber \\[4pt]
  & &  \quad\null
      +  \alpha {M\over \gs} [\gv\Vzero(\vecx)]^2 \phi(\vecx)
	  - \frac12 \alpha' [\gv\Vzero(\vecx)]^2 [\phi(\vecx)]^2
  \biggr) \ , \label{eq:efunctional}
\end{eqnarray}
subject to the constraint
\begin{equation}
  \int\! {\rm d}^3x \, \Ualpha^\dagger(\vecx) \Ualpha(\vecx) = 1 \ ,
  \label{eq:constraint}
\end{equation}
for all occupied states.
Observe that the functional has been truncated at quartic terms in the
fields and at quadratic terms in gradients of the fields; this truncation
will be justified below.

To realistically describe finite nuclei, the functional must be extended
to include rho mesons, pions, and photons.
In the present discussion, however,
we focus on nuclear ground-state properties that primarily
constrain isoscalar physics.
Thus the Coulomb contribution and a correct reproduction of the
bulk symmetry energy ($\approx 35\MeV$) are sufficient for our purposes,
and this requires only
the conventional extensions to include neutral rho mesons (denoted by $\bzero$)
and the Coulomb
field (see also Ref.~\cite{SEROT86}):
\begin{eqnarray}
     E[\{\Ualpha\},\phi,\Vzero,\bzero,\Azero] &=&
   E[\{\Ualpha\},\phi,\Vzero] +  \nonumber \\
     & & \qquad\null
    \int\! {\rm d}^3x \, \biggl(\,
  \sum_{\nu}^{\rm occ} \Ualpha^\dagger(\vecx) \bigl\{
   \frac12\grho\tau_3  \bzero(\vecx) + e\frac12(1+\tau_3)\Azero(\vecx) \bigl\}
     \Ualpha (\vecx)  \nonumber \\
    & & \qquad\qquad\quad\null
   - \frac12 \bigl\{ [\vecnabla\bzero(\vecx)]^2 + \mrho^2[\bzero(\vecx)]^2
            \bigr\}
      - \frac12 [\vecnabla\Azero(\vecx)]^2
  \biggr)
      \ .  \label{eq:isovector}
\end{eqnarray}
For simplicity, we suppress these isovector terms in the sequel,
although they are included in the numerical calculations of finite nuclei.
As noted above, the chiral models of Secs.~\ref{sec:sigma} and
\ref{sec:tang} have no explicit pion field contribution in this formulation.

The meson fields are determined by requiring the functional to be
stationary with respect to their variations; this yields (after
partial integrations) the meson field equations
\begin{eqnarray}
  (\vecnabla^2 - \ms^2) \phi(\vecx) &=&
    -\gs \sum_{\nu}^{\rm occ} \overline\Ualpha(\vecx)\Ualpha(\vecx)
      \nonumber \\  & & \qquad\null
    + \frac12 \kappa [\phi(\vecx)]^2
            + \frac{1}{6}\lambda [\phi(\vecx)]^3
      +  \Bigl(\alpha {M\over \gs} -  \alpha'\phi(\vecx)\Bigr)
             [\gv\Vzero(\vecx)]^2   \ ,  \label{eq:scalareq}  \\
  (\vecnabla^2 - \mv^2) \Vzero(\vecx) &=&
    -\gv \sum_{\nu}^{\rm occ} \Ualpha^\dagger(\vecx)\Ualpha(\vecx)
      \nonumber \\  & & \qquad\null
            + \frac{1}{6}\zeta \gv^4 [\Vzero(\vecx)]^3
      -  2\alpha {M\over \gs}\gv^2\Vzero(\vecx)\phi(\vecx)
       +  \alpha'
             \gv^2\Vzero(\vecx)[\phi(\vecx)]^2  \ .  \label{eq:vectoreq}
\end{eqnarray}
The single-particle orbitals are determined similarly.
The constraint equation (\ref{eq:constraint}) is imposed for
each orbital $\Ualpha$
by using a Lagrange multiplier $\epsilon_\nu$,
which we identify as the energy
eigenvalue of the Dirac equation:
\begin{equation}
\bigl\{
  -i\vecalpha\veccdot\vecnabla + \beta[M - \gs\phi(\vecx)]
   + \gv\Vzero(\vecx) \bigr\}  \Ualpha(\vecx) =
   \epsilon_\nu \Ualpha(\vecx) \ .  \label{eq:diraceq}
\end{equation}
The energy is minimized by solving these equations
self-consistently for the $N$ and $Z$ lowest eigenvalues
to determine the occupied neutron and proton states.

This energy functional, which provides an approximation to a general density
functional, could also be derived as the one-baryon-loop energy from the
lagrangian density
\begin{eqnarray}
	{\cal L} & = & \psibar [i\gamma_\mu \partial^\mu
	  - \gv \gamma_\mu V^\mu - (M - \gs \phi) ] \psi
	  +  \frac12(\partial_\mu \phi \partial^\mu \phi - \ms^2 \phi^2)
	\nonumber \\
	 &  & \quad\null
	    - \frac14 (\partial_\mu V_\nu - \partial_\nu V_\mu)^2
	    + \frac12 \mv^2 V_\mu V^\mu
	    - \frac{1}{3!}\kappa \phi^3
            - \frac{1}{4!}\lambda \phi^4
        \nonumber \\
	 &  & \quad\null  + \frac{1}{4!}\zeta \gv^4 (V_\mu V^\mu)^2
	  - \alpha \gv^2 {M\over \gs} V_\mu V^\mu \phi
	  + \frac12\alpha' \gv^2 V_\mu V^\mu \phi^2 \ ,
	\label{eq:bds1}
\end{eqnarray}
by following the discussion in Ref.~\cite{FURNSTAHL95}.%
\footnote{Note that this derivation of the functional starts from
a {\it Lorentz-invariant\/} action $S\!=\!\int {\rm d}^4 {\kern-.1em}x
\,{\cal L}$.
If we consider the energy functional as an effective functional, however,
we presently know of no reason to exclude terms that explicitly
contain the medium four-velocity $u^\mu$,
such as $u^\mu V_\mu V^\nu V_\nu$.
This issue will be considered in a later publication.\/}
Alternatively, one could
construct the Hartree Hamiltonian from Eq.~(\ref{eq:bds1})
and take its expectation
value in a state specified by static meson fields and a set of occupied
single-nucleon orbitals (labeled by quantum numbers $\nu$).
In this lagrangian formulation,
it appears that vacuum contributions are neglected,
because the sum in $E$ runs over valence nucleons only.
However, the vacuum effects can be precisely absorbed
into the coefficients of the interaction terms, provided we keep
nonlinearities to all orders \cite{FURNSTAHL95}.
If we assume that a truncation at some low order of the fields and their
derivatives is sufficiently accurate at nuclear densities
(which we justify on the basis of naturalness),
then the vacuum effects are already included
with sufficient accuracy in the functional of Eq.~(\ref{eq:efunctional}).

The notation and structure of the lagrangian
(\ref{eq:bds1}) has been chosen to
conform to previous usage in the literature.
It includes as special cases most of the lagrangians used in mean-field
studies.
The Walecka model retains the terms with nucleon fields and just
the kinetic and mass terms for the scalar and vector fields.
Nonlinear terms in the scalar field were introduced long ago by
Schiff \cite{SCHIFF51} and later studied by Boguta and Bodmer
\cite{BOGUTA77}.
Although widely used,
a long-standing worry about such models was that the quartic scalar
coupling $\lambda$ is negative in good fits
to finite nuclei \cite{REINHARD86,REINHARD89,GAMBHIR90,SHARMA93a}.

This concern is eliminated when quartic vector self-interactions are
included in the lagrangian or energy functional.
Such a term was first introduced by Bodmer and
Price \cite{BODMER89,BODMER91},
motivated by a search for softer high-density equations of state.
Gmuca employed the quartic vector term to account for density dependence
in the vector self-energy, as implied by DBHF calculations \cite{GMUCA92}.
Previous studies of lagrangians with neutral vector mesons have suggested
that such terms can lead to field equations with causality-violating
solutions \cite{VELO69,VELO73}
or to the mixing of spin-one and spin-zero representations \cite{OGIEVETSKI63}.
However, if $\zeta>0$ and the other vector parameters are natural, then
there are no problems with causality.
The second point is not an issue in our effective-field-theory approach; we are
not describing elementary vector mesons, and we certainly expect mixing
effects at finite density \cite{CHIN77}.

Equation~(\ref{eq:bds1}) also includes as special cases
all of the chiral models considered
in Ref.~\cite{FURNSTAHL93} and the chiral model of Ref.~\cite{FURNSTAHL95}.%
\footnote{The model of Ref.~\cite{FURNSTAHL95} actually contains
a logarithmic potential for the scalar
field, which means an infinite polynomial in $\phi$.  However,
for reasons discussed below, specifying the cubic and quartic terms
determines the energy functional at ordinary densities at a level
sufficient for our purposes.}
The relationships between parameters in Eq.~(\ref{eq:bds1})
and models from the literature are summarized in
Table~\ref{tab:one}.

\mediumtext
\begin{table}[tb]
\caption{Relationship between parameters of
Eq.~(\protect\ref{eq:efunctional})
or Eq.~(\protect\ref{eq:bds1}) and specific models.
Chiral models with a linear representation should have $\gs=\gpi$ at the level
considered here;
however, we will allow $\gs=\gpi/\gA$ with $\gpi\approx 13.5$
and $\gA \approx 1.26$,
so that the Goldberger--Treiman
relation is satisfied.
We have also taken $\mpi=0$ in writing the entries.
Note that the $\eta$'s in the last two
rows are unrelated.}
\smallskip
\begin{tabular}{llccccccc}
 Model  & Ref.\ & $\gv$  & $\kappa$ & $\lambda$ & $\zeta$ &
      $\alpha$ & $\alpha'$
     \\ \hline
 Walecka & \protect\cite{WALECKA74} & $\gv$ & 0 & 0 & 0 & 0 & 0 \\[4pt]
 Nonlinear scalar & \protect\cite{BOGUTA77} & $\gv$ & $-2b$ &
      $6c$ & 0 & 0 & 0 \\[4pt]
 Bodmer & \protect\cite{BODMER91} & $\gv$ & $2a$ &
      $6b$ & ${\textstyle 6\mv^2\over \textstyle Z^2\gv^2}$ & 0 & 0 \\[4pt]
 Chiral $\sigma\omega$ & \protect\cite{SEROT86} &
   $\gv$ & $-{\textstyle 3\gs \ms^2 \over \textstyle M}$ &
   ${\textstyle 3\gs^2\ms^2\over \textstyle M^2} $ &
     0 & 0 & 0 \\[4pt]
Boguta & \protect\cite{BOGUTA83} &
   ${\textstyle \gs\mv\over \textstyle M}$ &
    $-{\textstyle 3\gs \ms^2 \over \textstyle M}$
         & ${\textstyle 3\gs^2\ms^2\over \textstyle M^2} $ &
     0 & 1 & 1 \\[4pt]
General chiral & \protect\cite{FURNSTAHL93} &
   $\gv$ & $-{\textstyle 3\gs \ms^2 \over \textstyle M}$ &
   ${\textstyle 3\gs^2\ms^2\over \textstyle M^2} $ &
     $\zeta$ & $\eta^2$ & $\eta^2$  \\[4pt]
Nonlinear chiral & \protect\cite{FURNSTAHL95} &
 $ \gv$ &   ${\textstyle (3d-8)\ms^2 \over \textstyle  d\Szero}$ &
     ${\textstyle (11d^2 - 48d + 48)\ms^2 \over \textstyle (d\Szero)^2 }$ &
     $\zeta$ &
     ${\textstyle -\eta\gs \over \textstyle 2\Szero M}
     {\textstyle \mv^2\over\textstyle\gv^2 } $ & 0 \\[4pt]
\end{tabular}
\label{tab:one}
\end{table}

While our immediate intent is to work with a functional that includes
the chiral models as special cases,
the proposed functional (\ref{eq:efunctional})
is actually more general than it appears at first.
One can imagine several classes of additional terms in Eq.~(\ref{eq:bds1}):
\begin{enumerate}
\item Higher-order self-couplings and derivatives involving
meson fields alone [{\em e.g.}, $\phi^5$ or $(V^\mu V_\mu)^3$ or
  $(\partial_\mu \phi \partial^\mu \phi)^2$];
\item Contact terms involving nucleon fields beyond bilinear order
      [{\em e.g.}, $(\psibar\psi)^2$];
\item More complicated meson--nucleon couplings [{\em e.g.},
      $\gs(\phi)\psibar\psi\phi$].
\end{enumerate}
Let us consider each in turn.

For applications of the energy functional to ordinary nuclei,
contributions to the energy from higher-order polynomials (beyond quartic)
or higher-order gradient terms (beyond quadratic) are numerically small,
unless the coefficients are ``unnaturally'' large.
The definition of naturalness proposed here (and discussed further in
Sec.~\ref{sec:discuss}) is simple: when written in appropriate dimensionless
form, the coefficients of all mesonic terms in $E$ are of order unity.
Thus we can organize our truncation by counting powers of the
local scalar and vector potentials (or self-energies) that appear in the
single-particle Dirac equations, divided by the nucleon mass, and by
counting gradients of the potentials divided by the square of the mass.
We take as a basic principle that our functional {\it implicitly\/}
includes higher-order
terms, but with natural coefficients, which can be ignored
in practice at ordinary nuclear densities and below;
their small contributions can be absorbed into slight adjustments of the
other coefficients.
Thus we have the possibility of a useful expansion and truncation scheme in
this framework.
We must establish,
however, after determining the parameters, that the highest-order terms
retained do not dominate the energy, and that adding additional terms
produces only small changes in the parameters.
These considerations and the  consequences of this approach for
extrapolations to high density are discussed in Sec.~\ref{sec:discuss}.

Contact terms arise if we ``integrate out'' the scalar and vector fields
(by using the field equations, for example).
Conversely, contact terms included originally in the energy functional
(excluding certain terms with derivatives) can be eliminated in favor of
the scalar and vector fields, if we allow products of fields to all orders.
The issue then really becomes one of efficiency, since in either framework
one will have to truncate in practice.
Based on the economical successes of relativistic meson-exchange
phenomenology and to connect to specific models, we use the field expansion.
We again rely on the naturalness
of the coefficients to motivate a truncation at fourth order in the fields.%
\footnote{Strictly speaking, in this approach $\mv$ should
be a fit parameter rather than fixed at the physical $\omega$ meson mass.
We rely on resonance dominance and the insensitivity of the functional
to the precise value of this mass to justify fixing it.}
Further discussion and development of these ideas will be given in
a future paper \cite{FURNSTAHL96}.

Models have been proposed that replace $\Mstar = M - \gs\phi$
in the lagrangian by a more general function $\Mstar(\phi)$
[or equivalently, that make the replacement
$\gs \longrightarrow \gs(\phi)$] \cite{ZIMANYI90,DELFINO94,LENSKE95}.
This function must have a Taylor expansion around vanishing $\phi$, but
is otherwise quite general in principle.
More generally, the Yukawa couplings of the scalar and vector to the
nucleon fields can be replaced by polynomials
in $\phi$ and $V^\mu$ (in Lorentz-invariant combinations).
The motivation for such models is often that these terms are needed to
reflect the compositeness of the nucleon.
However, if we allow arbitrary polynomials in the meson fields in our
functional, we can simply {\em redefine\/}
the fields to eliminate non-Yukawa couplings to the nucleon.
(For example, we can replace $\phi \rightarrow \widetilde\phi$
where $\Mstar(\phi) \equiv M - \gs'\widetilde\phi$.)
The new functional has the same form as Eq.~(\ref{eq:efunctional}),
including all orders in the fields alone.
The observables are unchanged by the transformation, and the only question
is again one of efficiency, that is, whether a truncation to fourth order
in the fields is an accurate numerical approximation.

Thus, by working with Eq.~(\ref{eq:efunctional}), we accommodate a wide class
of effective models.
To start our analysis, we identify the nuclear observables that we wish our
functional to reproduce.
The properties of finite nuclei fall into three major categories:

\begin{enumerate}
  \item Nuclear shapes.  This includes the basic manifestations of
    nuclear saturation \cite{BOHR69,FETTER71}:
    a flat interior, a surface thickness independent of the baryon number
    $B$, and a systematic increase of the nuclear radius with $B^{1/3}$.
    These features are seen experimentally in the charge radius and
    the charge density; the latter is most clearly analyzed for this
    purpose in momentum space ({\em i.e.}, the form factor).
  \item Binding-energy systematics.  The binding energies of (closed-shell)
    nuclei across the periodic table must reflect the liquid-drop
    systematics of the basic semi-empirical mass formula.
    At a coarse level,
    this involves the sensitive interplay of the bulk binding energy
    ($a_1$), the surface energy ($a_2$), the Coulomb energy ($a_3$), and
    the bulk symmetry energy ($a_4$) (with a surface correction included)
    \cite{BOHR69,FETTER71,SEMF35,SEEGAR75}:%
    \footnote{There are much more sophisticated mass formulas incorporating
    additional physics, but this is the appropriate level of
    detail for our purpose.}
    \begin{equation}
      E/B - M = -a_1 + a_2 B^{-1/3} + a_3 {Z^2\over B^{4/3}}
            + a_4 {(N-Z)^2\over B^2 (1+3.28/B^{1/3})}
       \ .  \label{eq:semf}
    \end{equation}
  \item Single-particle properties.
    The single-particle potential is reflected in the ordering and
    spacing of single-particle levels, which are reasonably well identified
    experimentally (up to rearrangement effects that are comparatively small).
    Speaking in nonrelativistic language (since the Dirac equation can always
    be recast in ``Schr\"odinger-equivalent'' form),
    the level ordering in $\ell$ implies a central potential
    that interpolates between a harmonic oscillator and square well.
    In addition, the spin-orbit potential must be strong enough
    to ensure the correct
    shell closures, but not too strong, or else the systematics of
    nuclear deformations will not be reproduced \cite{FURNSTAHL87}.
    Finally, there is the energy dependence of the optical potential at
    low energies, which is related to the vector self-energy at the
    relativistic mean-field level \cite{SEROT86}.
\end{enumerate}
\noindent
One could add other properties to this list.
In many cases, however, these additional properties
are strongly correlated with the features listed above and so are reproduced
without imposing further conditions.

Using properties of nuclei as constraints, however, is difficult in the
sort of direct analysis we seek.
We will therefore extrapolate from the systematics of actual observables
(such as binding energies, rms radii, and spin-orbit splittings) and impose
conditions on the functional {\it in nuclear matter\/} near the equilibrium
density.
First, nuclear matter must exhibit a particular equilibrium density
($\rhozero$) and binding energy ($\ezero$) within a fairly narrow range.%
\footnote{We will use the convention that properties and fields evaluated
at equilibrium density are denoted with a subscript ``0''.}
These conditions are supported by calculations in relativistic models that
are fit directly to properties of finite nuclei, which
consistently predict similar $\rhozero$ and $\ezero$
when extrapolated to infinite nuclear matter \cite{REINHARD89,SHARMA93a}.

We can also consider derivatives of the energy with respect to the density
evaluated at equilibrium.
The second derivative or
curvature determines the compression modulus ($\Kzero$).
Even though the range in $\Kzero$ consistent
with nuclear properties is fairly broad (approximately 200--300 MeV),
this provides significant constraints on mean-field models.
The most direct effect of the compressibility on ground-state
properties is through the surface energy and is therefore manifested
in the energy systematics.
The surface energy is also correlated with the scalar mass.

The importance of higher derivatives of the energy is unclear at present.
Recent work has suggested that the ratio of the ``skewness''
(related to the third derivative) to the compression modulus is well
determined by nuclear monopole vibrations \cite{PEARSON91,RUDAZ92a}.
We do not use skewness to constrain the energy functional in this
investigation, but we comment on the correlation between skewness
and ground-state properties in Sec.~\ref{sec:discuss}.
Note that one can easily extend the analysis described
in Sec.~\ref{sec:Bodmer} to exploit additional constraints of this type.

To correctly reproduce empirical single-particle energies, the most
important ingredient beyond the depth and shape of the effective central
potential is the strength of the spin-orbit interaction.
In relativistic models,
the spin-orbit splittings are highly correlated
with the nucleon effective mass $\Mstar$.
An upper bound on $\Mstar$ is given by the reproduction of the shell
closures in heavy nuclei and the splittings of spin-orbit pairs in light
nuclei, while the reproduction of deformation systematics gives a lower
bound \cite{FURNSTAHL87}.
In conjunction with the other constraints,%
\footnote{Spin-orbit strength is not the only source of
constraints on $\Mstar$. Reinhard {\em et al}.\ concluded that
$\Mstarzero/M$
must be roughly 0.6 from detailed fits to finite nuclei that
{\it did not\/} include the spin-orbit splittings as input
\cite{REINHARD86}.\/}
$\Mstarzero/M$ must
be in a fairly narrow interval around 0.6.

The symmetry energy is determined by the observed energy systematics
[through Eq.~(\ref{eq:semf})], the relative energies of proton and
neutron single-particle levels in heavy nuclei \cite{HOROWITZ81},
and isotope shifts in neutron-rich nuclei \cite{SHARMA93a,SHARMA95}.
These imply that $a_4 \approx 35\MeV$, with an uncertainty of several MeV.

Finally, there are important restrictions on the ranges of the attractive
and repulsive interactions, or equivalently, on the momentum dependence of
the scalar and vector fields.
These conditions do not impact nuclear matter
(since the momentum transfer is zero at the mean-field level)
but are critical in finite nuclei.
The major constraint is on the scalar mass, which is consistently found
from direct fits to be
in the vicinity of 500 MeV (when the vector mass is taken to be the
physical $\omega$ mass).
Significantly larger scalar masses produce unobserved oscillations in the
charge density and too steep a surface in the effective central potential,
which leads to incorrect level orderings.
We will assess the importance of this constraint for chiral models
(and the success of the nuclear matter constraints)
by calculating finite
nuclei after the nuclear matter analysis.

One may ask: at what level of accuracy should we require nuclear observables
to be reproduced?
The answer depends on the planned application of the model and the goals of
the analysis.
For comparison with nuclear scattering experiments, accurate nuclear
densities and wave functions are required, but total binding energies are not
particularly important; in contrast, for studying fission barriers or the
properties of nuclei near the ``drip lines'', energy systematics are
paramount.
Since our goal is to exclude some classes of mean-field models,
we will impose less stringent constraints.
The nuclear matter equilibrium density $\rhozero$
and binding energy $\ezero$ are rather well known,
and small variations are not important to our conclusions.
We therefore fix
\begin{eqnarray}
  \rhozero &\equiv& \rhoBzero = 0.1484\fm^{-3} \qquad
   (\kfermi^0 = 1.30\fm^{-1}) \ ,  \label{eq:bds54} \\[4pt]
  \ezero &\equiv& {\edzero\over\rhozero} - M
     = -16.1\MeV = -a_1 \ .
   \label{eq:bds55}
\end{eqnarray}
We choose the ranges of acceptable $\Kzero$ and
$\Mstarzero$ based on models that have been directly optimized
for finite-nucleus observables, allowing considerable uncertainties.
Specifically, we consider
\begin{equation}
	180\MeV <  \Kzero  < 360 \MeV \ ,
	\label{eq:Kzero}
\end{equation}
\begin{equation}
	0.58 <  \Mstarzero/M  < 0.64  \ .
	\label{eq:Mstarzero}
\end{equation}
If one focuses on a subset of the nuclear observables,
one can stray outside these bounds and still obtain good results.
For example, the original Walecka model generates quite acceptable nuclear
densities despite its  high compression modulus, which yields
unfavorable energy systematics \cite{HOROWITZ81}.


\section{Constraining Nuclear Matter Energy Functionals}
\label{sec:Bodmer}

Given the conditions on the energy functional for nuclear matter
[namely, $\ezero$, $\rhozero$, $\Kzero$, and $\Mstarzero$
in Eqs.~(\ref{eq:bds54})--(\ref{eq:Mstarzero})],
our goal is to solve for the parameters of the functional
as functions of these input ``observables.''
In general, we can determine four of the parameters from
these four inputs, and the additional coefficients
will then  specify the space of acceptable models.
To rule out a class of models, as we will do in Sec.~\ref{sec:sigma},
we must show that it is impossible to find a set of coefficients that fall
within the bounds of acceptability.

In uniform nuclear matter, the  energy functional of
Eq.~(\ref{eq:efunctional}) simplifies to the energy density function
\begin{eqnarray}
	\ed(\phi,\Vzero;\rhoB) & = & \gv \Vzero \rhoB - \frac12\mv^2\Vzero^2
	   + \gv^2\Vzero^2 \Big(\alpha {M\over \gs}\phi
	            - \frac12\alpha' \phi^2\Big)
 - \frac{1}{4!}\zeta \gv^4 \Vzero^4	\nonumber \\[4pt]
	 &  & \quad\null  + \frac12 \ms^2 \phi^2
	   + {\kappa\over 6}\phi^3
  +  {\lambda\over 24 }\phi^4 +
	   {\gamma\over (2\pi)^3}\int^{\kfermi}\! {\rm d}^3p\
	   (\pvec^2 + \Mstar{}^2)^{1/2}  \ ,
	\label{eq:bds2}
\end{eqnarray}
where
\begin{equation}
	\Mstar \equiv M - \gs \phi \ ,
	\label{eq:Mstar}
\end{equation}
and the Fermi momentum is defined through
\begin{equation}
  \rhoB \equiv {\gamma\kfermi^3\over 6 \pi^2} \ .
     \label{eq:kfermi}
\end{equation}
The spin-isospin degeneracy $\gamma$ is 4 for symmetric nuclear matter.
The coefficients $\kappa$ and $\lambda$ may be constrained by chiral
symmetry (as in the linear sigma model), but are free at this point.
Note also, that if we choose to work with a renormalizable lagrangian
and include the one-baryon-loop vacuum energy $\Deltaevac$
(either chirally or not), this is equivalent to redefining  $\kappa$
and $\lambda$ as well as the coefficients of higher powers of $\phi$
(which are not shown).
As before, the parameters and their special values in particular models
are given in Table~\ref{tab:one}.

It is useful to change variables in the energy density.
First we switch from $\phi$ and $\Vzero$ to the mean-field nucleon
self-energies (or optical potentials) $\Phi$ and $W$
(following the notation of Ref.~\cite{BODMER91}):
\begin{equation}
   \Phi \equiv \gs \phi \ ,
  \qquad\qquad
    W \equiv \gv \Vzero  \ .
	\label{eq:bds9}
\end{equation}
The meson field equations are now determined by
extremizing $\ed$ with respect to $\Phi$ and $W$.
As discussed earlier, the natural expansion parameters
for the finite-density functional are $W/M$ and $\Phi/M$.
The quadratic terms in Eq.~(\ref{eq:bds2}) imply that
the couplings $\gs$ and $\gv$ do not appear individually in nuclear
matter, so we rewrite the energy density in terms of variables that
absorb the couplings.
We introduce the ratios of meson couplings to masses
\begin{equation}
  \cs \equiv {\gs \over \ms} \ , \qquad\qquad	\cv \equiv {\gv \over \mv}\ ,
	\label{eq:bds10}
\end{equation}
together with the scalar self-couplings
\begin{equation}
    \kappabar \equiv {\kappa\over\gs^3} \ ,\qquad\qquad
    \lambdabar \equiv {\lambda\over\gs^4}  \ ,  \label{eq:klbar}
\end{equation}
and the scalar--vector couplings
\begin{equation}
  \alphabar \equiv {\alpha\over \gs^2} \ , \qquad\qquad
     \alphabarp \equiv {\alpha'\over \gs^2}  \ . \label{eq:scavec}
\end{equation}
The correspondences between conventional model parameters and the newly
defined parameters are summarized in Table~\ref{tab:two}.
The nucleon mass $M = 939\MeV$ and the vector meson mass $\mv =
m_{\omega} = 783\MeV$
will be fixed at their experimental values.

\mediumtext
\begin{table}[t]
\caption{Relationship between parameters of Eq.~(\protect\ref{eq:bds2})
and specific models.  The quadratic coupling is $\cs = \gs/\ms$.
Chiral models with a linear representation should have $\gs=\gpi$ at the level
considered here;
however, we will allow $\gs=\gpi/\gA$ with $\gpi\approx 13.5$ and
$\gA \approx 1.26$,
so that the Goldberger--Treiman relation is satisfied.
We have also taken $\mpi=0$ in writing the entries.
Recall that the $\eta$'s in the last two rows are unrelated.}
\smallskip
\begin{tabular}[t]{llccccccc}
 Model  & Ref.\ & $\cv$  & $\kappabar$ & $\lambdabar$ & $\zeta$ &
      $\alphabar$ & $\alphabarp$
     \\ \hline
 Walecka & \protect\cite{WALECKA74} & $\textstyle\gv\over\textstyle\mv$
             & 0 & 0 & 0 & 0 & 0 \\[4pt]
 Nonlinear scalar & \protect\cite{BOGUTA77} & $\textstyle\gv\over\textstyle\mv$
      & $-{\textstyle 2b\over\textstyle\gs^3}$ &
      $\textstyle 6c\over\textstyle\gs^4$ & 0 & 0 & 0 \\[4pt]
 Bodmer & \protect\cite{BODMER91} & $\textstyle\gv\over\textstyle\mv$
      & $\textstyle 2a\over\textstyle\gs^3$ &
      $\textstyle 6b\over\textstyle\gs^4$ &
      $\textstyle6\mv^2\over \textstyle Z^2\gv^2$
                     & 0 & 0 \\[4pt]
 Chiral $\sigma\omega$ & \protect\cite{SEROT86} &
   $\textstyle\gv\over\textstyle\mv$ &
   $-{\textstyle 3\over\textstyle\cs^2 M}$ &
    ${\textstyle 3\over \textstyle\cs^2  M^2} $ &
     0 & 0 & 0 \\[4pt]
Boguta & \protect\cite{BOGUTA83} &
   ${\textstyle\gs\over \textstyle M}$ &
   $-{\textstyle 3\over \textstyle\cs^2 M}$
         & ${\textstyle 3\over \textstyle\cs^2 M^2} $ &
     0 & $\textstyle 1\over\textstyle\gs^2$ &
     $\textstyle 1\over\textstyle \gs^2$ \\[4pt]
General chiral & \protect\cite{FURNSTAHL93} &
   $\textstyle\gv\over\textstyle\mv$ &
   $-{\textstyle 3 \over \textstyle\cs^2 M}$ &
    ${\textstyle 3\over \textstyle\cs^2  M^2} $ &
     $\zeta$ & $\textstyle \eta^2\over\textstyle\gs^2$
     & $\textstyle\eta^2\over\textstyle\gs^2$  \\[4pt]
Nonlinear chiral & \protect\cite{FURNSTAHL95} &
 $ \textstyle\gv\over\textstyle\mv$  &
   ${\textstyle(3d-8) \over \textstyle d(\gs\Szero)\cs^2}$ &
     ${\textstyle(11d^2 - 48d + 48) \over \textstyle(d\gs\Szero)^2\cs^2 }$ &
      $\zeta$ &
     ${\textstyle -\eta \over \textstyle 2(\gs\Szero) \cv^2 M} $ & 0 \\[4pt]
\end{tabular}
\label{tab:two}
\end{table}

In terms of the new variables, the energy density can be written as
\begin{equation}
  \ed(\Phi,W;\rhoB) = \edvphi(\Phi,W;\rhoB) + U(\Phi) + \edk(\Phi;\rhoB) \ ,
	\label{eq:bds15}
\end{equation}
where
\begin{equation}
	\edvphi(\Phi,W;\rhoB) = W\rhoB - \frac{1}{2\cv^2} W^2
	+ W^2\Big(\alphabar M\Phi - \frac12\alphabarp \Phi^2 \Big)
	  - \frac{\zeta}{24} W^4
	   \ ,
	\label{eq:bds16}
\end{equation}
\begin{equation}
  U(\Phi) = {1\over 2\cs^2}\Phi^2 + {\kappabar\over 6}\Phi^3
      + {\lambdabar\over 24 }\Phi^4 \ ,
	\label{eq:bds17}
\end{equation}
and $\edk$ is the ``kinetic energy'' (including the nucleon rest mass):
\begin{eqnarray}
	\edk(\Phi;\rhoB) & = & {\gamma\over (2\pi)^3} \int^{\kfermi}\!
         {\rm d}^3p\, \sqrt{\pvec^2+\Mstar{}^2}
	\nonumber \\[4pt]
	 & = & {\gamma\over 8\pi^2} \biggl[
	   \kfermi \Efermistar{}^3 - \frac12 \Mstar{}^2\kfermi\Efermistar
	 - \frac12\Mstar{}^4\ln\biggl({\kfermi+\Efermistar\over\Mstar}\biggr)
	      \biggr]  \ ,
	\label{eq:bds18}
\end{eqnarray}
with
\begin{equation}
	\Efermistar \equiv \sqrt{\kfermi^2 + \Mstar{}^2} \ .
	\label{eq:bds18a}
\end{equation}
This division is useful because it
isolates the $W$ dependence and  the  parameters in $U$ ($\cs^2$, $\kappabar$,
and $\lambdabar$),
for which we will solve.
We could also solve for a different set, as in Sec.~\ref{sec:sigma}.

The energy per particle ($\ed/\rhoB-M$) and pressure $p$ are determined by
the seven free parameters $\cv$, $\cs$, $\kappabar$, $\lambdabar$,
$\alphabar$, $\alphabarp$, and $\zeta$.
Any two models with the same values of these constants will give identical
results for nuclear matter, so these are the relevant parameters for
testing our constraints, rather than those in the original energy
functional.

Let us summarize the conditions on $\ed$ as a function of $\Phi$,
$W$, and $\rhoB$ \cite{BODMER91}.
By evaluating these conditions at equilibrium, we will be  able to solve
for the parameters in terms of the input data $\ezero$, $\rhozero$, $\Kzero$,
and $\Mstarzero$.
We observe immediately that if $\Mstarzero$ is given, then $\Phizero$
follows trivially from
  \begin{equation}
     \Phizero = M - \Mstarzero  \ .
     \label{eq:phizero}
  \end{equation}
It is therefore convenient to define the equilibrium values
\begin{equation}
  \Uzero \equiv U(\Phizero)
  = {1\over 2\cs^2}\Phizero^2 + {\kappabar\over 6}\Phizero^3
      + {\lambdabar\over 24 }\Phizero^4
  \ ,  \label{eq:Uzero}
\end{equation}
\begin{equation}
  \Uzerop \equiv U'(\Phizero)={\d U(\Phi)\over \d\Phi}\biggr|_{\Phizero}
  = {1\over \cs^2}\Phizero + {\kappabar\over 2}\Phizero^2
      + {\lambdabar\over 6 }\Phizero^3
  \ ,  \label{eq:Uzerop}
\end{equation}
and
\begin{equation}
  \Uzeropp \equiv U''(\Phizero) =
    {\d^2 U(\Phi) \over \d\Phi^2}\biggr|_{\Phizero}
  = {1\over \cs^2} + \kappabar\Phizero + {\lambdabar\over 2 }\Phizero^2
  \ .      \label{eq:uzeroppdef}
\end{equation}
The conditions on $\ed$ are as follows:
\begin{enumerate}
  \item At any density $\rhoB$, $\ed$ is stationary with respect to
  $\Phi$:
  \begin{equation}
      \biggl({\partial\ed\over\partial \Phi}\biggr)_{W,\rhoB} = 0 \ .
        \label{eq:Wcons}
  \end{equation}
  Carrying out this derivative on Eq.~(\ref{eq:bds15}) and evaluating
  the result at equilibrium yields
  \begin{equation}
   \rhoszero - \Wzero^2[\alphabar M - \alphabarp \Phizero]
      - \Uzerop = 0
       \ ,  \label{eq:uzerop}
  \end{equation}
  where
\begin{equation}
  \rhoszero \equiv \rhos(\Phizero,\rhozero) \ ,
     \label{eq:rhoszerodef}
\end{equation}
with
  \begin{eqnarray}
    \rhos(\Phi,\rhoB) =
          -\biggl({\partial\edk\over\partial \Phi}\biggr)_{W,\rhoB}
        &=& {\gamma\over (2\pi)^3} \int^{\kfermi}\! {\rm d}^3 p\,
	   {\Mstar \over\sqrt{\pvec^2+\Mstar{}^2}}  \nonumber \\[4pt]
      &=&  {\gamma\Mstar\over 4\pi^2}\biggl[\kfermi\Efermistar - \Mstar{}^2
                 \ln\biggl({\kfermi+\Efermistar\over\Mstar}\biggr)\biggr]
                  \ .  \label{eq:rhosdef}
  \end{eqnarray}
  \item
The vector field appears only in $\edvphi$ and is determined by the equation
of constraint
\begin{equation}
	\biggl({\partial\ed\over\partial W}\biggr)_{\Phi,\rhoB}
  = \biggl( {\partial \edvphi\over \partial W}\biggr)_{\Phi,\rhoB} = 0 \ ,
	\label{eq:bds19}
\end{equation}
which produces (at equilibrium)
\begin{equation}
     \rhozero - {1\over \cv^2}\Wzero
        + \Wzero[2\alphabar M\Phizero - \alphabarp\Phizero^2]
       - {\zeta\over 6} \Wzero^3 = 0
	  \ .
	\label{eq:bds20}
\end{equation}
  \item
  The binding energy per particle at $\rhoB=\rhozero$ is $-\ezero$, so
  \begin{equation}
     \edzero = (\ezero + M)\rhozero   \ .  \label{eq:ev}
  \end{equation}
  \item
The Hugenholtz--van Hove theorem follows by
calculating the chemical potential $\mu$ from $\ed$ in Eq.~(\ref{eq:bds15}):
\begin{equation}
  \mu =	{\d\ed\over \d\rhoB}
   = W + \biggl({\partial\edk\over\partial \rhoB}\biggr)_{\Phi}
   = W + \sqrt{\kfermi^2 + \Mstar{}^2} \ .
	\label{eq:bds41}
\end{equation}
Note that we need only take $(\partial\ed/\partial\rhoB)_{\Phi,W}$
to obtain this
result, since $\ed$ is stationary with respect to $\Phi$ and $W$
from Eqs.~(\ref{eq:Wcons}) and (\ref{eq:bds19}).
The mean-field energy density is thermodynamically consistent,
so the pressure
follows from the first law of thermodynamics (as can be explicitly verified
by taking derivatives):
\begin{equation}
	p = - \ed + \rhoB \mu = -\ed + \rhoB W + \rhoB \Efermistar \ .
	\label{eq:bds42}
\end{equation}
The condition for equilibrium is $p=0$ at $\rhoB \neq 0$,
so combining
Eq.~(\ref{eq:ev}) with Eq.~(\ref{eq:bds42})
evaluated at equilibrium yields
\begin{equation}
  \Wzero =
  \ezero + M - \sqrt{(\kfermizero)^2 + \Mstarzero{}^2}
  \ .   \label{eq:bds56}
\end{equation}
This quite general result provides a direct connection between $\Wzero$ and
the inputs $\ezero$, $\rhozero$, and $\Mstarzero$ \cite{BOGUTA77,BODMER91}.
  \item
The compression modulus $K$ follows from
\begin{equation}
  {1 \over 9} K(\rhoB) \equiv \rhoB^2 {\d^2(\ed/\rhoB)\over \d\rhoB^2}
      = \rhoB {\d^2\ed \over \d\rho_B^2} - 2 {p\over\rhoB}
  \ , \label{eq:bds52}
\end{equation}
so that at equilibrium we have
\begin{eqnarray}
  \Kzero \equiv K(\rhozero) &=& 9\rhozero{\d^2\ed\over \d\rhoB^2}
                     \biggr|_{\rhozero} \nonumber \\
        &=& 9\rhozero \biggl[
          {\pi^2\over 2\kfermizero\Efermistarzero} +
          {\partial W \over \partial \rhoB}\biggr|_{\Phizero,\rhozero}
          +\biggl({\partial W\over \partial\Phi}\biggr|_{\Phizero,\rhozero}
            - {\Mstarzero\over \Efermistarzero} \biggr)
            {\d\Phi\over \d\rhoB}\biggr|_{\rhozero}
          \biggr]
                      \ .
\end{eqnarray}
Carrying out the derivatives using the field equations%
\footnote{This is most simply done by considering $W$ to be a function of
$\Phi$ and $\rhoB$.}
produces
\begin{equation}
   \Kzero = 9\rhozero \left[
     {\pi^2\over 2 \kfermizero\Efermistarzero} + \mzero
     - { \Bigl({\textstyle\Mstarzero\over\textstyle\Efermistarzero}
      - \mzero\lzero\Bigr)^2 \over
         \Uzeropp
         - \rhospzero
         + \lzero^2\mzero - \alphabarp\Wzero^2 }
      \right]
   \ ,  \label{eq:bds51}
\end{equation}
where
\begin{equation}
  \lzero \equiv {2\over \Wzero}(\rhoszero - \Uzerop)
    = 2 \Wzero (\alphabar M - \alphabarp \Phizero) \ ,
 \label{eq:lzero}
\end{equation}
\begin{equation}
  \mzero \equiv {\Wzero \over \rhozero + \frac13\zeta\Wzero^3}
    = {1 \over {1/ \cv^2} - (2\alphabar\Phizero M - \alphabarp\Phizero^2)
       + \frac12\zeta \Wzero^2 }  \ ,
 \label{eq:mzero}
\end{equation}
and
\begin{equation}
  \rhospzero \equiv {\partial\rhos \over \partial\Phi}\biggr|_{\Phizero}
     = -{\partial\rhos \over \partial\Mstar}\biggr|_{\Mstarzero}
 \ ,  \label{eq:rhospzerodef}
\end{equation}
with
\begin{equation}
  {\partial\rhos \over \partial\Mstar} =
   {\rhos\over \Mstar} + {2\over\pi^2} \Mstar{}^2 \biggl[
        {\kfermi\over \Efermistar} -
    \ln \biggl( {\kfermi+\Efermistar\over \Mstar} \biggr) \biggr]
   \ . \label{eq:bds49}
\end{equation}
\end{enumerate}

We can now systematically solve for $\cv^2$ and the three parameters
in $U(\Phi)$ using the four inputs $\ezero$, $\rhozero$, $\Kzero$,
and $\Mstarzero$.
This leaves as free parameters $\zeta$, $\alphabar$, and $\alphabarp$,
which label potentially viable solutions.
We note immediately that $\Mstarzero$ gives us $\Phizero$ from
Eq.~(\ref{eq:phizero}) and then $\Wzero$ from Eq.~(\ref{eq:bds56}).
We can now solve Eq.~(\ref{eq:bds20}) for $\cv^2$, since given $\zeta$,
$\alphabar$, and $\alphabarp$, everything else is known:
\begin{equation}
  \cv^2 = {\Wzero \over \rhozero + \Wzero [2 \alphabar M \Phizero
            -\alphabarp\Phizero^2] - \frac16\zeta\Wzero^3}
            \ .
  \label{eq:cveq}
\end{equation}
Next, we express $\cs^2$, $\kappabar$,
and $\lambdabar$ in terms of $\Uzero$, $\Uzerop$, and $\Uzeropp$ by
inverting Eqs.~(\ref{eq:Uzero})--(\ref{eq:uzeroppdef}):%
\footnote{At this stage, the form of $U(\Phi)$ is not critical,
and if $U(\Phi)$ has three unknown parameters (but not
necessarily a quartic polynomial), one can just calculate $\Uzero$,
$\Uzerop$, and $\Uzeropp$
and invert to find the parameters in terms of $\rhozero$, $\ezero$,
$\Mstarzero$, and $\Kzero$ \protect\cite{BODMER91}.}
\begin{eqnarray}
  \cs^2 &=& {\frac{1}{2}\Phizero^2 \over
            \frac12\Phizero^2\Uzeropp - 3 \Phizero \Uzerop + 6\Uzero}  \ ,
  \label{eq:cssq}   \\[8pt]
  \kappabar &=& {-\Phizero^2\Uzeropp +5 \Phizero \Uzerop - 8\Uzero
           \over  \frac{1}{6}\Phizero^3}  \ ,
  \label{eq:kappabar}  \\[8pt]
  \lambdabar &=& {\frac12\Phizero^2\Uzeropp - 2 \Phizero \Uzerop + 3\Uzero
           \over  \frac{1}{24}\Phizero^4}  \ .
  \label{eq:lambdabar}
\end{eqnarray}

The final step is to relate $\Uzero$, $\Uzerop$, and $\Uzeropp$
to the input observables.
After determining $\Phizero$, $\Wzero$, and $\cv$,  we can calculate
$\edvphi$ and $\edk$ at equilibrium.
Then $\Uzero \equiv U(\Phizero)$ follows from Eqs.~(\ref{eq:bds15})
and (\ref{eq:bds55}):
\begin{eqnarray}
 \Uzero &=& \edzero - \edvphi(\Phizero,\Wzero;\rhozero) -
 \edk(\Phizero;\rhozero)
            \nonumber \\
   &=& (\ezero+M)\rhozero
         - \edvphi(\Phizero,\Wzero;\rhozero) - \edk(\Phizero;\rhozero) \ .
  \label{eq:bds64}
\end{eqnarray}
Here $\ezero$, $\rhozero$, $\Mstarzero$, and $\zeta$ are given,
$\edk(\Phizero;\rhozero)$ follows from Eq.~(\ref{eq:bds18}) at the
given $\rhozero$ and $\Mstarzero$, and we use
Eqs.~(\ref{eq:bds16}) and (\ref{eq:bds20})
to get $\edvphi(\Phizero,\Wzero;\rhozero)$, with the result
\begin{equation}
  \edvphi(\Phizero,\Wzero;\rhozero) = \frac12 \Wzero\rhozero
     + \frac{1}{24} \zeta \Wzero^4  \ .
  \label{eq:newevphi}
\end{equation}
We can evaluate $\Uzerop$ using Eqs.~(\ref{eq:uzerop}) and
(\ref{eq:rhosdef}):
\begin{equation}
     \Uzerop =
    \rhoszero - \Wzero^2[\alphabar M - \alphabarp \Phizero]
    \label{eq:auzerop}
\end{equation}
and then solve the $\Kzero$ equation (\ref{eq:bds51})
for $\Uzeropp$ in terms of known quantities:
\begin{equation}
  \Uzeropp =  \rhospzero
    - \lzero^2\mzero  + \alphabarp\Wzero^2  +
    { \Bigl({\textstyle\Mstarzero\over\textstyle\Efermistarzero}
      - \mzero\lzero\Bigr)^2 \over
    {\textstyle\pi^2\over \textstyle 2\kfermizero\Efermistarzero}
       + \mzero - {\textstyle\Kzero\over \textstyle 9\rhozero} }
     \ .  \label{eq:uzeropp}
\end{equation}

If we use all four inputs $\ezero$, $\rhozero$, $\Mstarzero$, and $\Kzero$
to determine the parameters $\cv^2$, $\cs^2$,
$\kappabar$, and $\lambdabar$, we can find solutions
parametrized by the remaining free couplings, which are
$\zeta$, $\alphabar$, and $\alphabarp$ in the general case.
It would seem that we can always find a solution.
How, then, can we ever exclude a model?
There are several possibilities.
First, the model may be less general, so that there are not enough
free parameters.
For example, the Walecka model has only two free parameters ($\cs^2$ and
$\cv^2$); thus, only $\ezero$ and $\rhozero$ can be specified, and
$\Mstarzero$ and $\Kzero$ are predicted.
There is a transcendental equation for $\Mstarzero$ to be solved
in this case [see Eq.~(\ref{eq:mstartrans}), below],
and the solution may not fall within the bounds specified by
Eq.~(\ref{eq:Mstarzero}).
Another possibility is that the solution manifests
unphysical conditions ({\em e.g.}, $\cs^2<0$ or $\cv^2<0$).

A third possibility is that the parameters allow multiple
solutions to the meson field equations,  which may
correspond to pathological or abnormal solutions
({\em e.g.}, a Lee--Wick solution \cite{LEE74} in the sigma model).
These alternative solutions extremize the energy functional and may
have a lower energy than the normal solution (which satisfies the
phenomenological input conditions).
As discussed in the next section, we will not necessarily exclude
a model in this case unless the normal solution, which exists
by construction at $\rhozero$, disappears at a slightly higher density.
If this occurs, we must consider the parameter set unacceptable.


\begin{figure}
\setlength{\epsfxsize}{4in}
\centerline{\epsffile{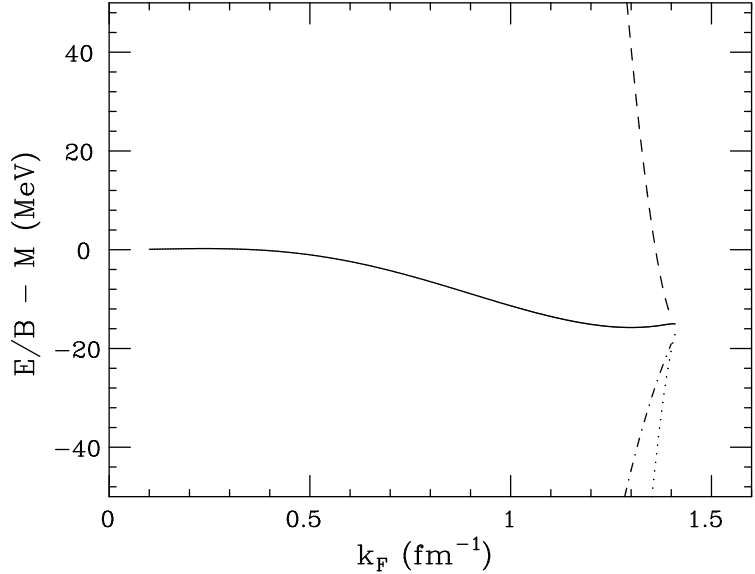}}
\vspace*{.1in}

\caption{\protect\footnotesize%
Binding-energy curves with $\zeta = -0.03$,
$\alphabar=\alphabarp=0$, and all other parameters determined
from $\ezero$, $\rhozero$, $\Mstarzero$, and $\Kzero$.
All of the curves extremize the energy
functional.  The normal solution (with equilibrium binding energy
$-16.1\MeV$ at $\kfermi=1.3\infm$) is the solid line.
Only four of the six solutions are shown.%
}
\label{fig:ten}
\end{figure}


To investigate the latter possibility, we must always
check for multiple solutions to the $\Phi$ and $W$ equations
[(\ref{eq:uzerop}) and (\ref{eq:bds20})] after determining the parameters.
Equation~(\ref{eq:bds20}) is easy to analyze, since it is
a cubic in $W$ with known coefficients (for given $\Phi$).
For example, one can show that with $\alphabar=\alphabarp=0$,
it is necessary that $\zeta>0$ to avoid abnormal nuclear matter
solutions with lower energy than the normal solution.
We illustrate what can happen with multiple solutions in Fig.~\ref{fig:ten},
for $\alphabar=\alphabarp=0$ and $\zeta=-0.03$.
The solid line is the normal solution, with a minimum at $\ezero$ and
$\rhozero$.
The other curves arise from different solutions to the $W$ and $\Phi$
equations (three each!).
The normal solution disappears at a density slightly above
$\rhozero$,  and only pathological
solutions remain at higher densities (not shown).

A final possibility is that the parameter set is unnatural, that is,
the coefficients of the highest-order terms are too large.
These terms then dominate the energy functional
at densities near nuclear matter equilibrium, which implies that our
truncation is unjustified, and the entire framework breaks down.

The Bodmer method for determining the model parameters from information
about nuclear matter equilibrium can be conveniently implemented in a
spreadsheet or symbolic manipulation program.
It can be summarized as follows:
\def\labelenumi{\theenumi.}
\renewcommand{\theenumi}{\arabic{enumi}}
\begin{enumerate}
  \item Specify $\ezero$, $\rhozero$, $\Kzero$, $\Mstarzero$,
          and also $\zeta$, $\alphabar$, and $\alphabarp$.
          (Here we have chosen to solve for $\cv^2$, $\cs^2$,
          $\kappabar$, and $\lambdabar$.)
  \item $\Phizero$ follows from Eq.~(\ref{eq:phizero}) and $\Wzero$ from
        Eq.~(\ref{eq:bds56}).
  \item Determine $\cv^2$
         from Eq.~(\ref{eq:cveq}). If $\mv$ is specified, then
         $\gv^2$ follows immediately from the definition
         in Eq.~(\ref{eq:bds10}).
  \item Compute $\Uzero$ from Eqs.~(\ref{eq:bds64}), (\ref{eq:newevphi}), and
        (\ref{eq:bds18}); $\Uzerop$ from Eqs.~(\ref{eq:rhosdef})
         and~(\ref{eq:auzerop}); and
        $\Uzeropp$ from Eqs.~(\ref{eq:lzero})--(\ref{eq:bds49})
        and~(\ref{eq:uzeropp}).
        Solve for $\cs^2$, $\kappabar$, and $\lambdabar$ from
        Eqs.~(\ref{eq:cssq})--(\ref{eq:lambdabar}).
        If the scalar mass $\ms$ is also specified,
        one can obtain $\gs$, $\kappa$, and $\lambda$ from
        Eqs.~(\ref{eq:bds10}) and (\ref{eq:klbar}).
  \item Find all solutions to the scalar and vector equations.
     (It is easiest to first eliminate $W$ from the $\Phi$ equation
     by directly solving the cubic $W$ equation.)
     Check that the normal solution is acceptable within a prescribed
     density range.
\end{enumerate}

As a starting point for our discussion, we set $\alphabar = \alphabarp = 0$,
so that we consider the same class of models studied by
Bodmer \cite{BODMER91}.
A useful procedure is to fix $\ezero$ and $\rhozero$ and then to scan
the ($\Kzero$, $\Mstarzero$) plane, with each point producing a family
of parameter sets, since there are five free parameters and only four
constraints.
We can then establish boundaries corresponding to special values of the
parameters.

\begin{figure}
\setlength{\epsfxsize}{4.5in}
\centerline{\epsffile{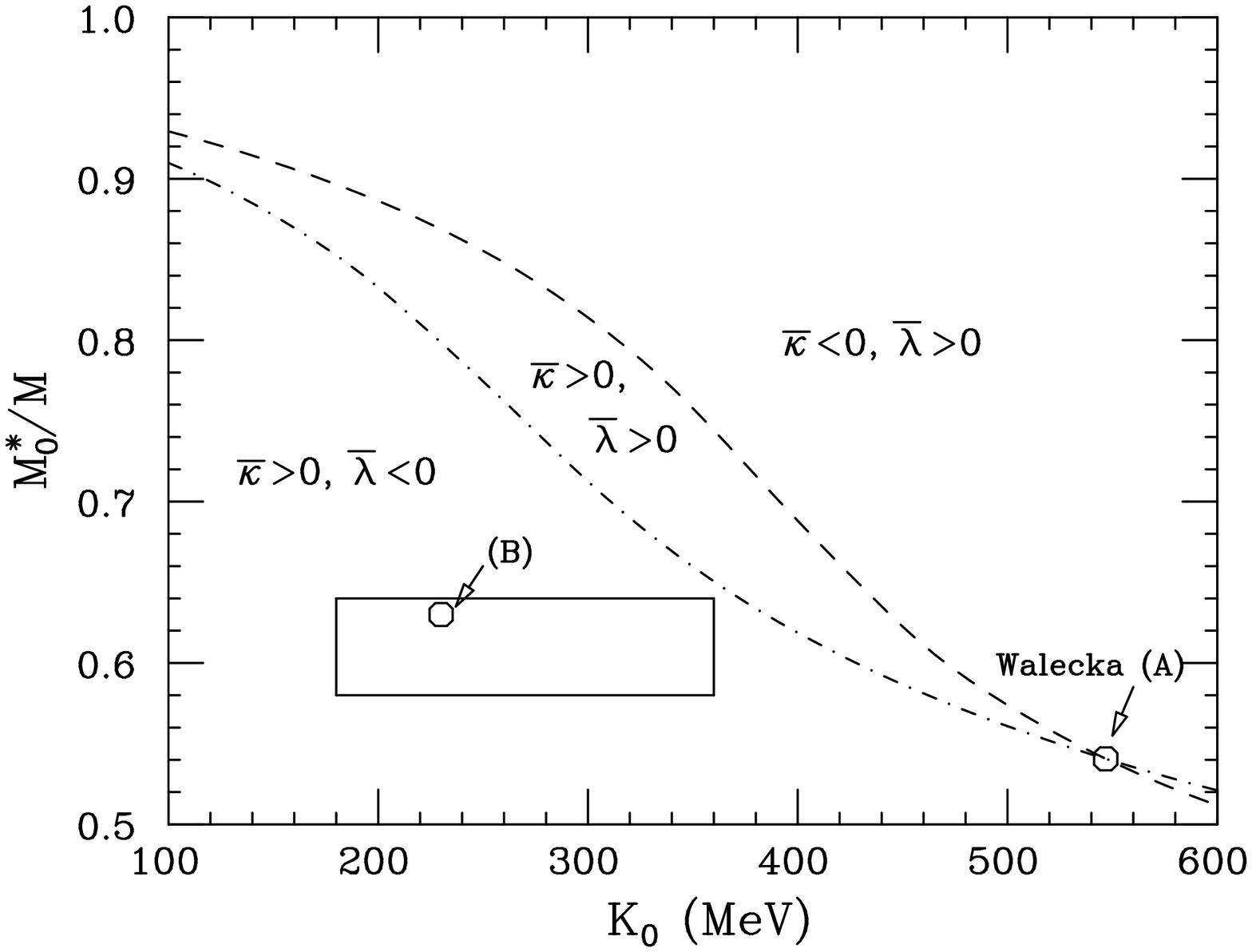}}
\vspace*{.1in}
\caption{\capcrunch{\protect\footnotesize%
 $\Mstarzero/M$ vs.\ $\Kzero$
 with $\zeta=\alphabar=\alphabarp=0$.
 The lines show the
 $\kappabar=0$ and $\lambdabar=0$ boundaries.
 The box encloses the desirable values of $\Mstarzero/M$ and $\Kzero$.
 Model~B is from Ref.~\protect\cite{FURNSTAHL93}
 (see Table~\protect\ref{tab:twop}).%
}}

\label{fig:one}
\end{figure}

With $\alphabar=\alphabarp=0$, the energy density reduces to
\begin{eqnarray}
  \ed &=& W\rhoB + \edk(\Phi; \rhoB)
    - {1\over 2\cv^2}W^2
	  + {1\over 2\cs^2}\Phi^2
+ \frac16\kappabar\Phi^3 + \frac{1}{24}\lambdabar \Phi^4
	   - {1\over 24}\zeta W^4
	  \ .  \label{eq:bds40p}
\end{eqnarray}
This implies that
\begin{itemize}
	\item  $\zeta > 0$ lowers the energy (attractive),

	\item  $\kappabar$, $\lambdabar>0$ raise the energy (repulsive),

	\item  $\kappabar$, $\lambdabar<0$ lower the energy (attractive).
\end{itemize}
If $\kappabar = \lambdabar = 0$ (no scalar nonlinearities) and $\zeta=0$
(no vector nonlinearities), then we obtain the original Walecka model.
In equilibrium, we can set $\ed$ from Eq.~(\ref{eq:bds40p}) equal to
$(\ezero+M)\rhozero$, and eliminate $\Wzero$ using Eq.~(\ref{eq:bds56}),
$\cs^2$ using Eq.~(\ref{eq:uzerop}), and $\cv^2$ using Eq.~(\ref{eq:bds20}).
This leaves a transcendental equation for $\Mstarzero$:
\begin{equation}
     (\ezero+M)\rhozero + \sqrt{(\kfermizero)^2 + \Mstarzero{}^2}\, \rhozero
          - (M-\Mstarzero)\rhoszero - 2 \edk(\Mstarzero;\rhozero) = 0
     \ . \label{eq:mstartrans}
\end{equation}
For the values of $\rhozero$ and $\ezero$ in Eqs.~(\ref{eq:bds54}) and
(\ref{eq:bds55}), the solution is $\Mstarzero/M = 0.538$, which then
implies $\Kzero = 557\MeV$ from Eq.~(\ref{eq:bds51}).
This point is denoted as (A) in Fig.~\ref{fig:one}.

If $\zeta=0$ but $\kappabar\neq0$ and $\lambdabar\neq 0$, it is possible
to reproduce the desired $\Mstarzero$ and $\Kzero$ with $U(\Phi)$ alone,
which is well known.
In this case, $\kappabar>0$ and $\lambdabar<0$.
The situation can be displayed graphically in the following figures.
In Fig.~\ref{fig:one}, we plot the locus of points for which $\kappabar =0$
or $\lambdabar = 0$, as a function of both $\Mstarzero/M$ and $\Kzero$.
This lets us separate the regions of parameter space corresponding
to different signs of $\kappabar$ and $\lambdabar$.
The boxed area shows the acceptable values of the input ``observables''
($0.58 \leq \Mstarzero/M \leq 0.64$,\ \ $180 \leq \Kzero \leq 360\MeV$).
Note that when $\lambdabar<0$, the energy is unbounded below
for large $\Phi$;
this is not necessarily a problem,
however, since our truncated functional is only valid for small $\Phi$.

\begin{figure}
\setlength{\epsfxsize}{4.5in}
\centerline{\epsffile{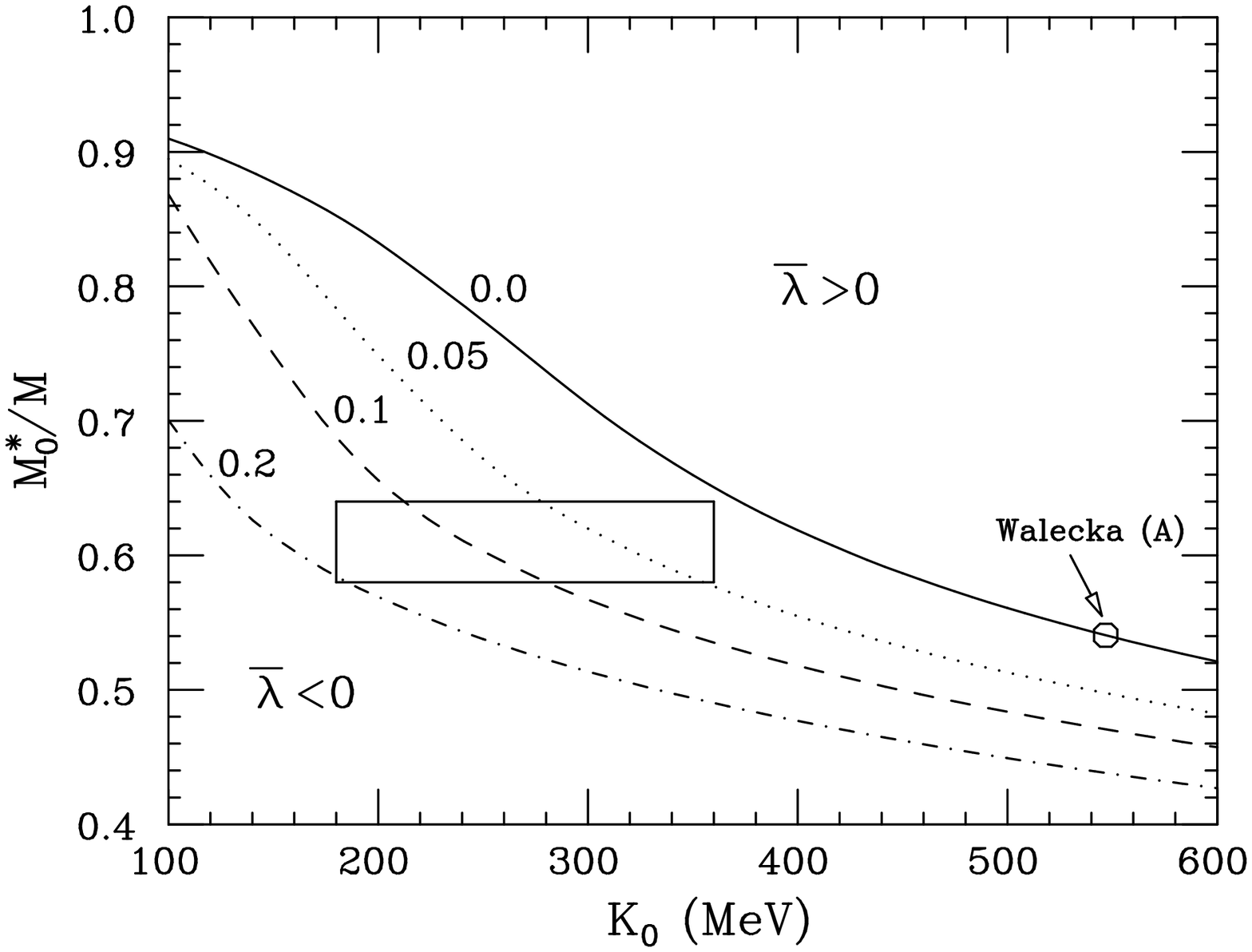}}
\vspace*{.1in}
\caption{\capcrunch{\protect\footnotesize%
 $\Mstarzero/M$ vs.\ $\Kzero$ with
 $\alphabar=\alphabarp=0$.
 The lines show the
 $\lambdabar=0$ boundaries for a range of values of
 $\cv^2\zeta \Wzero^2/6$ from 0.0 to 0.2.
 This combination is a measure of the vector nonlinearities.
 The box encloses the desirable values of $\Mstarzero/M$ and $\Kzero$.%
}}

\label{fig:two}
\end{figure}


If we now allow $\zeta\neq 0$ (recall that $\zeta$ must be positive), then
as $\zeta$ increases, we get additional attractive contributions to the
energy, which leads us to increase $\lambdabar$.
Figure~\ref{fig:two} shows lines of $\lambdabar=0$ for selected values of
$\cv^2\zeta \Wzero^2/6$.
[This combination is the ratio of the cubic and linear terms in
Eq.~(\ref{eq:bds20}) and is thus the relevant measure of vector
nonlinearities.]
For large enough $\zeta$, we can achieve acceptable $\Mstarzero$
and $\Kzero$ with $\lambdabar>0$.
We can see how this works by examining the expression for $\ed$ in
Eq.~(\ref{eq:bds40p}).
Since both $\Phizero$ and $\Wzero$ are approximately proportional to
$\rhoB$, a phenomenologically acceptable $\ed$ clearly requires an explicit
dependence $\propto \rhoB^4$ with a negative coefficient.
If $\zeta=0$, we must set $\lambdabar < 0$ to achieve this.
By setting $\zeta > 0$, however, we can increase $\lambdabar$ and obtain
essentially the same result.
Nevertheless, we must be careful not to let the quartic nonlinearities get
so large that they dominate the energy, for this would violate our
assumption of naturalness.

\begin{figure}
\setlength{\epsfxsize}{4.5in}
\centerline{\epsffile{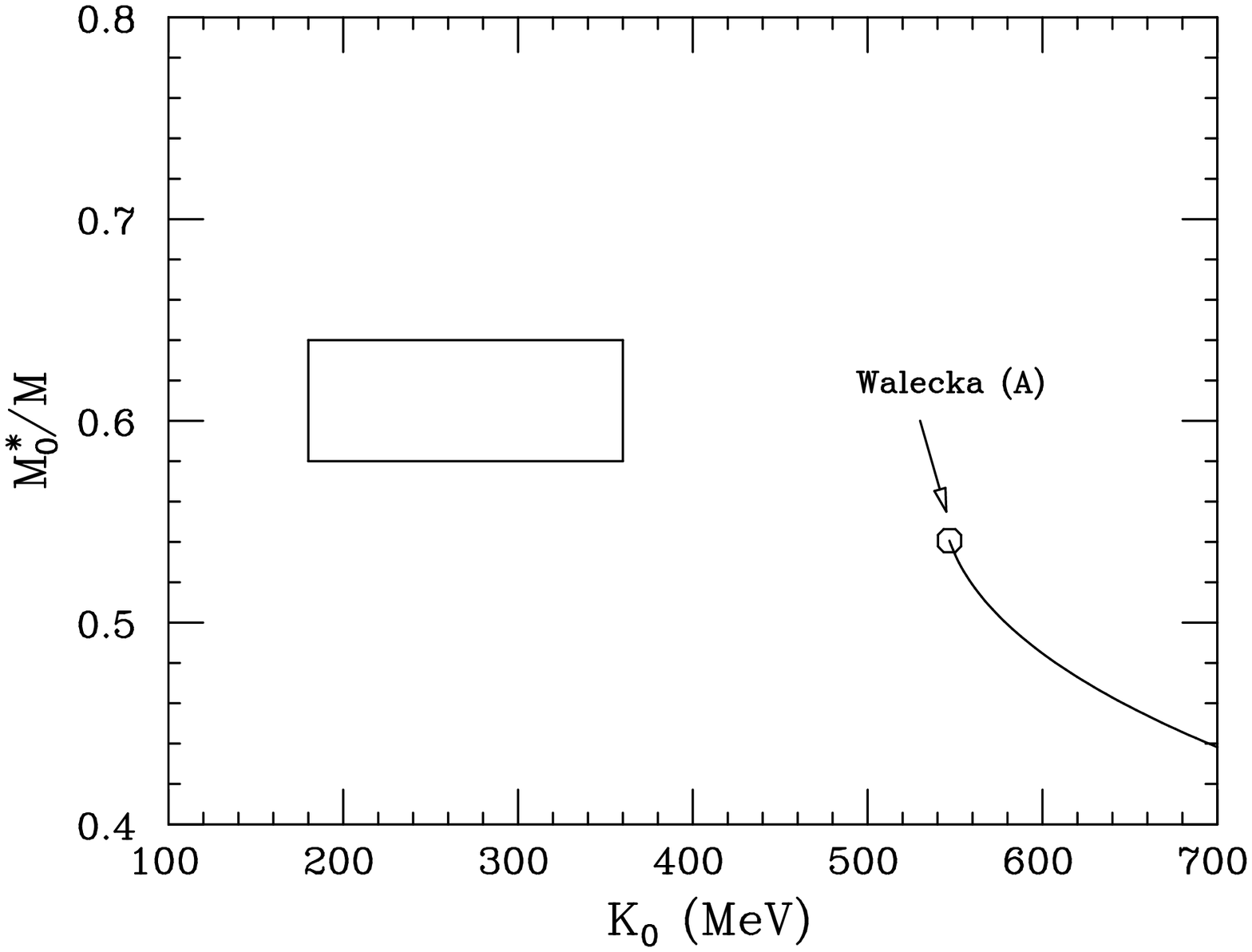}}
\vspace*{.1in}
\caption{\capcrunch{\protect\footnotesize%
 $\Mstarzero/M$ vs.\ $\Kzero$ with
 $\alphabar=\alphabarp=0$ and $\kappabar=\lambdabar=0$.
 The line shows possible solutions as a
 function of $\zeta \cv^2 \Wzero^2/6$.
 The $\zeta=0$ point is the Walecka model.
 The box encloses the desirable values of $\Mstarzero/M$ and $\Kzero$.%
}}

\label{fig:twop}
\end{figure}

If we hold $\kappabar = \lambdabar = \alphabar =\alphabarp =0$
and let $\zeta > 0$ (vector nonlinearities only), then $\Mstarzero/M$
decreases and $\Kzero$ increases relative to the Walecka model, as indicated
by the line in Fig.~\ref{fig:twop}.
Thus adding $\zeta$ alone makes the resulting nuclear matter properties
worse; to return to the desired region of $\Mstarzero$ and $\Kzero$, we
must restore the scalar nonlinearities.
Furthermore, when one increases $\zeta$, one must compensate by increasing
$\lambda$, so that an acceptable description is obtained only when both
terms are included.
This conclusion is consistent with our definition of naturalness and is
also supported by the density dependence of the scalar and vector
self-energies found in DBHF calculations \cite{GMUCA92}.


One can further explore the space of acceptable parametrizations
by letting $\alphabar \neq 0 $ and $\alphabarp \neq 0$.
Although the enlarged parameter space allows for a wide variety of acceptable
solutions, this additional freedom is not needed to satisfy the
phenomenological constraints considered here.
Nevertheless, two observations can be made.
First, including the $\alphabar$ and $\alphabarp$ parameters allows one to
adjust the density dependence of the effective scalar and vector meson masses
(defined by diagonalizing the matrix of appropriate second derivatives
of the energy functional),
which may provide useful constraints in the future if
concrete empirical information becomes available.
 (Note that one must
be careful of the distinction between longitudinal and transverse
masses when trying to compare with experiment.)
These additional parameters may also allow for improved fits to DBHF
self-energies or to self-energies obtained in more sophisticated nuclear
matter calculations.
Second, our definition of naturalness says that there is no reason to omit
these scalar--vector couplings, unless there is some (as yet unknown)
symmetry principle that forbids them.
In the interest of brevity, we leave an exploration of the expanded
parameter space to a future investigation.
We turn instead to specific chiral models, considered as special cases of
the full energy functional,
and apply a variant of the Bodmer analysis described above.

\section{Analysis of Linear Sigma Models}
\label{sec:sigma}

We now adapt the general analysis of the previous section
to consider mean-field models built on the linear sigma model with
a light scalar meson.
There is a long history of such models, starting with
the lagrangian considered by Kerman and Miller \cite{KERMAN74,SEROT86}:
\begin{eqnarray}
 {\cal L} &=& \psibar[i\gamma_\mu\partial^\mu - \gv\gamma_\mu V^\mu
    - \gpi(\sigma + i \gamma_5 \vectau\veccdot\vecpi)]
   \psi
   \nonumber \\[4pt]
  & & \null
  + \frac12 (\partial_\mu\sigma \partial^\mu\sigma
 +  \partial_\mu\vecpi {\bf \cdot}
                   \partial^\mu\vecpi)
  -  \frac14\lambda(\sigma^2 + \vecpi^2 - v^2)^2
  \nonumber \\[4pt]
  & & \null
  - {1\over 4} (\partial_\mu V_\nu - \partial_\nu V_\mu)^2
  + {1\over 2}\mv^2 V_\mu V^\mu  + \epsilon\sigma \ .  \label{eq:kmeq}
\end{eqnarray}
Because of the ``Mexican-hat'' potential, $v\neq 0$ implies that $\sigma$
acquires a nonzero vacuum expectation value, which generates masses for the
nucleon and scalar meson (and pion).
This also fixes the self-couplings of the scalar field $\phi$ that is
the deviation of $\sigma$ from its vacuum value.
Here we study descriptions of nuclei that arise from generalizations
of this model.
We stress that we are now exploring the phenomenological
consequences of definite lagrangians
treated in the Hartree approximation.
We are able to use the effective density functional analysis developed
in the previous sections because the form of the energy functional is
the same.

The general extension of the linear sigma model (\ref{eq:kmeq}) considered in
Ref.~\cite{FURNSTAHL93} is based on the lagrangian (after shifting the scalar
field $\sigma$ from its vacuum value to define $\phi$)
\begin{eqnarray}
 {\cal L} &=& \psibar[i\gamma_\mu\partial^\mu - \gv\gamma_\mu V^\mu
    - (M - \gpi\phi)   - i \gpi\gamma_5 \vectau\veccdot\vecpi]
   \psi
   \nonumber \\[4pt]
  & & \null
  + {1\over 2}(\partial_\mu\phi \partial^\mu\phi - \ms^2\phi^2 )
 + {1\over 2}(\partial_\mu\vecpi {\bf \cdot}
                   \partial^\mu\vecpi - \mpi^2 \vecpi^2)
  \nonumber \\[4pt]
  & & \null
  + {\gpi\over 2M}(\ms^2 - \mpi^2) \phi (\phi^2 + \vecpi^2)
 - {\gpi^2\over 8 M^2} (\ms^2 - \mpi^2)(\phi^2 + \vecpi^2)^2
  \nonumber \\[4pt]
  & & \null
  - {1\over 4} (\partial_\mu V_\nu - \partial_\nu V_\mu)^2
  + {1\over 2}\mv^2 V_\mu V^\mu
  + {1\over 4!}\zeta\gv^4(V_\mu V^\mu)^2
  \nonumber \\[4pt]
  & & \null
  - \eta^2 \left({\gv^2\over\gpi}\right) M V_\mu V^\mu \phi
 + {1\over 2}\eta^2 \gv^2 V_\mu V^\mu (\phi^2 + \vecpi^2)
  \ .  \label{eq:sigmashift}
\end{eqnarray}
Here we have eliminated $\lambda$, $v$, and $\epsilon$ in favor
of $M$, $\ms$, and $\mpi$.
We will take $\mpi=0$ in the sequel.%
\footnote{When solving chiral mean-field models, one must be aware
of the possibility of bound ``anomalous'' solutions, in which the
scalar field interpolates between the minima of the effective potential.
Including a finite pion mass increases the energy of the anomalous
solution so that it is above the energy of the normal solution.
Hence we can simply ignore these anomalous solutions and
set $\mpi=0$. (The pion mass has negligible effect on ``normal''
solutions.)}
This is the most general lagrangian with nonderivative couplings through
dimension four that is consistent with linear chiral
symmetry (for $\mpi=0$).

Note that chiral symmetry in the linear realization implies that
the scalar--nucleon Yukawa coupling constant $\gs$ is equal to
the pion--nucleon coupling constant $\gpi$, which is known experimentally.
Moreover, the nucleon mass satisfies $M = \gpi\fpi$ at the
one-loop level.
Since this implies $\gA=1$, in contradiction to experiment, we will
allow $\gs = \gpi/\gA$ when working at the Hartree level.

In the Hartree approximation, the pion field vanishes for nuclear
matter or axially symmetric nuclei, and the energy functional derived from
Eq.~(\ref{eq:sigmashift}) reduces to a special case of the functional
considered earlier.
In nuclear matter there are four free parameters:
$\cs^2$, $\cv^2$, $\alphabar=\alphabarp=\eta^2/\gs^2$, and $\zeta$.
We can adapt the analysis of Sec.~\ref{sec:Bodmer} to solve
for these parameters in terms of the inputs $\ezero$, $\rhozero$,
$\Kzero$, and $\Mstarzero$.
Naively, this would seem to imply that we {\it can\/} find good sigma-model
descriptions of finite nuclei, in contrast to the
conclusion of Ref.~\cite{FURNSTAHL93}.
We first present the analysis and then discuss why this is
not the case.

To specialize the general functional of Secs.~\ref{sec:setup}
and \ref{sec:Bodmer} to the chiral models
from Ref.~\cite{FURNSTAHL93}, we set [see Eq.~(\ref{eq:bds17})]
\begin{equation}
  \alphabarp = \alphabar = {\eta^2\over\gs^2} > 0  \ ,  \label{eq:alphabarch}
\end{equation}
\begin{equation}
   \kappabar = -{3\over M}{1\over \cs^2} < 0\ , \label{eq:kappabarch}
\end{equation}
\begin{equation}
   \lambdabar = {3\over M^2}{1\over \cs^2} > 0\ . \label{eq:lambdabarch}
\end{equation}
Now $\ed$ takes the form
  \begin{eqnarray}
     \ed &=& W \rhoB  + \edk(\Phi;\rhoB) -
     {1\over 2\cv^2}W^2  - \frac{1}{4!}\zeta W^4
     \nonumber  \\[4pt]  & & \qquad\null
         + {1\over 2\cs^2}\Phi^2
         \biggl(1-{\Phi\over M}+{\Phi^2\over 4 M^2}\biggr)
      + \frac12 \alphabar W^2 (M^2-\Mstar{}^2)
     \ .  \label{eq:evchiral}
  \end{eqnarray}
At equilibrium, $\Phizero$ still follows trivially from $\Mstarzero$:
\begin{equation}
     \Phizero = M - \Mstarzero  \ ,
     \label{eq:phizerochi}
\end{equation}
and the scalar potential and its derivatives can be evaluated from
Eqs.~(\ref{eq:Uzero})--(\ref{eq:uzeroppdef}) using
the relations (\ref{eq:kappabarch}) and (\ref{eq:lambdabarch}).
We can then simply rewrite the equilibrium conditions from
Sec.~\ref{sec:Bodmer}
(as before, a subscript ``0'' denotes the value at equilibrium):
\begin{enumerate}
  \item The scalar field equation becomes
  \begin{equation}
    \rhoszero - \alphabar \Mstarzero \Wzero^2
      - {1\over\cs^2}\Phizero \biggl[1 - {3\Phizero\over 2M}
          + {\Phizero^2\over 2M^2}  \biggr]
    =     \rhoszero - \alphabar \Mstarzero \Wzero^2
       - {\Mstarzero (M^2 - \Mstarzero{}^2)\over 2\cs^2 M^2} = 0
       \ ,  \label{eq:uzeropchi}
  \end{equation}
  where $\rhos$ is still given by Eq.~(\ref{eq:rhosdef}).
  \item
The vector field  constraint equation is
\begin{equation}
     \rhozero - {1\over \cv^2}\Wzero + \alphabar\Wzero[M^2 - \Mstarzero{}^2]
       - {\zeta\over 6} \Wzero^3 = 0
	  \ .
	\label{eq:bds20chi}
\end{equation}
  \item
  The energy density can be written explicitly as
  \begin{eqnarray}
     \edzero &=& (\ezero + M)\rhozero   \nonumber \\[4pt]
     &=& \frac12 \Wzero\rhozero + \frac{1}{24}\zeta \Wzero^4
         + {1\over 2\cs^2}\Phizero^2
         \biggl(1-{\Phizero\over M}+{\Phizero^2\over 4 M^2}\biggr)
           + \edk(\Phizero;\rhozero)
     \ ,  \label{eq:evchi}
  \end{eqnarray}
with $\edk(\Phizero;\rhozero)$ evaluated from Eq.~(\ref{eq:bds18}).
The second equality is obtained from Eq.~(\ref{eq:evchiral}) using the
equation of constraint (\ref{eq:bds20chi}).
Observe that there is no explicit dependence on $\alphabar$ in
Eq.~(\ref{eq:evchi}).
  \item
The Hugenholtz--van Hove theorem goes through as before, and we again find
\begin{equation}
  \Wzero =
  \ezero + M - \sqrt{(\kfermizero)^2 + \Mstarzero{}^2}
  \ .   \label{eq:bds56chi}
\end{equation}
  \item
The compression modulus at equilibrium is
\begin{equation}
   \Kzero = 9\rhozero \Biggl[
     {\pi^2\over 2 \kfermizero\Efermistarzero} + \mzero
     - { \Bigl({\Mstarzero/\Efermistarzero} - \mzero\lzero\Bigr)^2 \over
         \Uzeropp
         - \rhospzero
         + \lzero^2\mzero - \alphabar\Wzero^2 }
      \Biggr]
   \ ,  \label{eq:bds51chi}
\end{equation}
with $\rhospzero$ still evaluated
from Eq.~(\ref{eq:bds49}), but now
\begin{equation}
  \lzero
    = 2 \alphabar \Mstarzero\Wzero \ ,
 \label{eq:lzerochi}
\end{equation}
\begin{equation}
  \mzero = {\Wzero \over \rhozero + \frac13\zeta\Wzero^3} \ ,
 \label{eq:mzerochi}
\end{equation}
and
\begin{equation}
  \Uzeropp = {1\over \cs^2} \biggl[ 1 - 3{\Phizero\over M}
        + {3\Phizero^2\over 2M^2} \biggr] \ .
  \label{eq:uzeroppchi}
\end{equation}
\end{enumerate}

One can solve
Eqs.~(\ref{eq:phizerochi})--(\ref{eq:uzeroppchi}) for
$\cs^2$, $\alphabar$, $\zeta$, and $\cv^2$ with some straightforward algebra.
After eliminating $\alphabar$, $\zeta$, and $\cv^2$, the resulting equation
involving $\cs^2$ is actually linear and yields
\begin{equation}
\cs^2 = {\textstyle
     {\textstyle\Wzero \Mstarzero{}^2 \over \textstyle M^2}
     + {\textstyle\Mstarzero{}^2 \xi_4 \over \textstyle(\Efermistarzero)^2}
     - {\textstyle4 \Mstarzero \xi_5 \over \textstyle\mathstrut\Efermistarzero}
     - \xi_1 \left( {\textstyle\Mstarzero{}^2 \over \textstyle M^2}\,
     \xi_2 + \xi_3 \xi_4
          - {\textstyle8 \rhoszero \over \textstyle\mathstrut\Wzero}
          \xi_5 \right)
  \over  \textstyle
  \xi_1 \left( {\textstyle4 \rhoszero^2 \over \textstyle\mathstrut\Wzero}
       -\xi_2 \xi_3
        \right)
     + \Wzero \xi_3
      - {\textstyle4 \Mstarzero \rhoszero \over
             \textstyle\mathstrut\Efermistarzero}
    + {\textstyle\Mstarzero{}^2  \over \textstyle(\Efermistarzero)^2}\, \xi_2
     } \ , \label{eq:cschiral}
\end{equation}
where
\begin{eqnarray}
\xi_1 &\equiv& {\Kzero \over 9 \rhozero} - {\pi^2 \over 2 \kfermizero
    \Efermistarzero} \ , \label{eq:xi1}\\[4pt]
\xi_2 &\equiv& {8\over \Wzero}\, [(\ezero + M )\rhozero - \edkzero]
     - 3\rhozero \ , \label{eq:xi2}\\[4pt]
\xi_3 &\equiv& \rhospzero + \rhoszero/ \Mstarzero \ , \label{eq:xi3}\\[4pt]
\xi_4 &\equiv& {\Phizero^2 \over \Wzero}\,
      \left( 1 + {\Mstarzero \over M}\right)^2
     \ , \label{eq:xi4}\\[4pt]
\xi_5 &\equiv& \Phizero \left( 1 - {3 \Phizero \over 2 M}
    + {\Phizero^2 \over 2 M^2} \right)
      \ , \label{eq:xi5}
\end{eqnarray}
with $\edkzero \equiv \edk(\rhozero,\Mstarzero)$.
The remaining parameters are  determined from
\begin{eqnarray}
\alphabar &=& {1 \over \Wzero^2 \Mstarzero}\,
      \left[ \rhoszero - {1 \over \cs^2}\, \Phizero
      \left( 1 - {3 \Phizero \over 2 M}
           + {\Phizero^2 \over 2 M^2} \right)\right]
       \ , \label{eq:alphabarchi}\\[4pt]
\zeta &=& {24\over \Wzero^4}\,
      \left[ (\ezero + M -\frac12\Wzero )\rhozero
     - \edkzero - {1 \over 2 \cs^2}\, \Phizero^2
     \left( 1 - {\Phizero \over M} + {\Phizero^2 \over 4 M^2} \right)\right]
     \ , \label{eq:zetachi}\\[4pt]
\cv^2 &=& {\Wzero \over
     \rhozero + \alphabar \Wzero (M^2 - \Mstarzero{}^2) - \zeta
          \Wzero^3/6 }   \ . \label{eq:cvchi}
\end{eqnarray}

We can use these results to explore the entire range of
acceptable normal solutions.
The procedure to find parameter sets is direct:
specify $\ezero$, $\rhozero$, $\Mstarzero$,
and $\Kzero$, and the parameters follow algebraically.
Nevertheless, the resulting parameter sets will not always be acceptable.
For example, parametrizations with $\cs^2<0$ or $\cv^2<0$ are unphysical.
In most cases, however, the main issue is the influence of the Lee--Wick
solution.

If we recall from Eq.~(\ref{eq:rhosdef}) that $\rhos \propto \Mstar$,
it is clear that at any density,
Eq.~(\ref{eq:uzeropchi}) will {\it always\/} be satisfied
when $\Mstar=0$ or $\Phi=M$.
This is the Lee--Wick solution, which is an alternative to the ``normal''
solution, the latter being identified as having zero energy at
$\rhoB=0$.
(In general, there is also an ``anomalous'' solution, which arises
from a local {\em maximum\/} in the energy density as a function of
$\Mstar$.)
There is no evidence that Lee--Wick states exist in nature.
Notwithstanding, while the Lee--Wick solution exists formally at all
densities, we are not guaranteed that the normal solution specified at
equilibrium will persist at higher density (it may coalesce with the
``anomalous'' solution and disappear).


\begin{figure}
\setlength{\epsfxsize}{4.5in}
\centerline{\epsffile{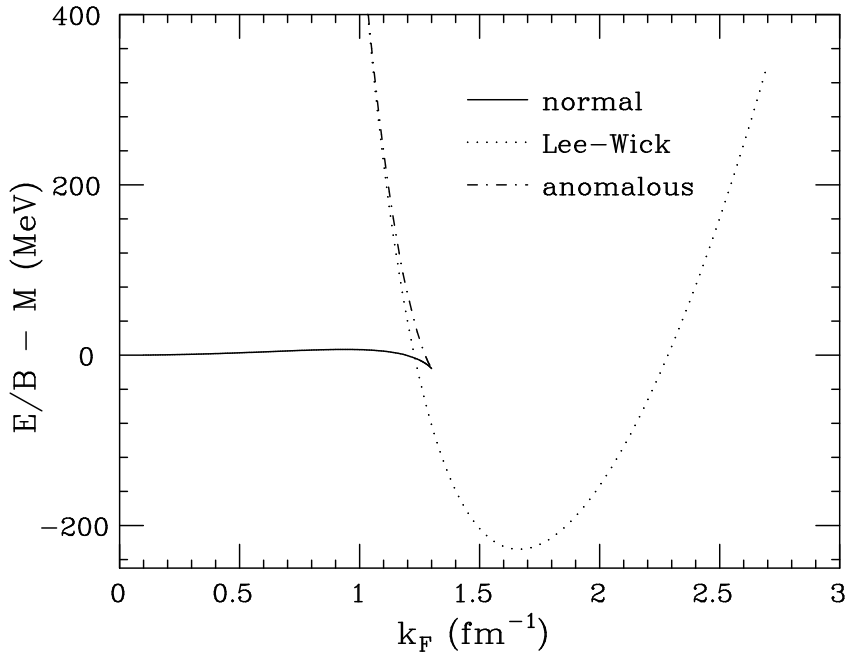}}
\vspace*{.1in}
\caption{\capcrunch{\protect\footnotesize%
 Nuclear matter binding-energy curves
for the simple linear $\sigma\omega$ model.
The normal solution has the standard $\ezero$ at density $\rhozero$, but
this is not a point of equilibrium.%
}}

\label{fig:eleven}
\end{figure}


An example of the difficulties of bifurcation and coalescence that
typically arise from the highly nonlinear field equations
is found in the original work of Kerman and Miller \cite{KERMAN74}.
In this $\sigma\omega$ model [Eq.~(\ref{eq:evchiral}) with
$\alphabar=\zeta=0$], there are only two free parameters, as in the
original Walecka model.
Proceeding as in that case (see Sec.~\ref{sec:Bodmer}), one discovers that
there is {\it no normal solution\/} that reproduces the desired $\ezero$
and $\rhozero$ as an equilibrium point.
The best one can do is to satisfy one of the conditions, for example,
to reproduce the desired energy/particle $\ezero$ at a density $\rhozero$;
however, the system will not be in equilibrium (the pressure is negative),
and moreover, the normal solution fails to exist at densities $\rho >
\rhozero$.
This situation is depicted in Fig.~\ref{fig:eleven}.
Note that the Lee--Wick solution has the lowest energy at most densities
and is the only solution at high density.

These results suggest that we might consider a parametrization acceptable
only if it produces a normal state with the lowest energy, at least up to
the equilibrium density.
Alternatively, if one recalls that our energy density is
truncated at some finite power of $\Phi /M$, one might say that the
Lee--Wick solution with $\Phi = M$ is simply irrelevant.
In particular, one could argue that it is possible to add (chirally
invariant) repulsive terms of
very high order in $\Phi / M$ that would have negligible effect on the
normal solution, but which increase the energy of the Lee--Wick state
enough to raise it above the normal state.
Rather than digress into a somewhat extraneous discussion about the proper
role of the Lee--Wick solution in an effective field theory, we will instead
impose on our parameter sets a less restrictive, more physical constraint
that is incontrovertible:
If a normal solution is found that satisfies the desired inputs, it
cannot {\it disappear\/} at too low a density to be useful.

To make the consequences of this constraint more concrete, so that we can
determine the boundaries of acceptable parametrizations,
we will fix $\ezero$ and $\rhozero$ as usual and then scan the ($\Kzero$,
$\Mstarzero$) plane, finding the unique normal parameter set
($\cs^2$, $\cv^2$, $\alphabar$, and $\zeta$) at each point.
This allows us to determine regions of the plane in which it is impossible
to find a normal solution at $\rhozero$, in which a normal solution exists
at $\rhozero$ but disappears at some higher density, and in which a normal
solution exists at all densities.


\begin{figure}
\setlength{\epsfxsize}{4.5in}
\centerline{\epsffile{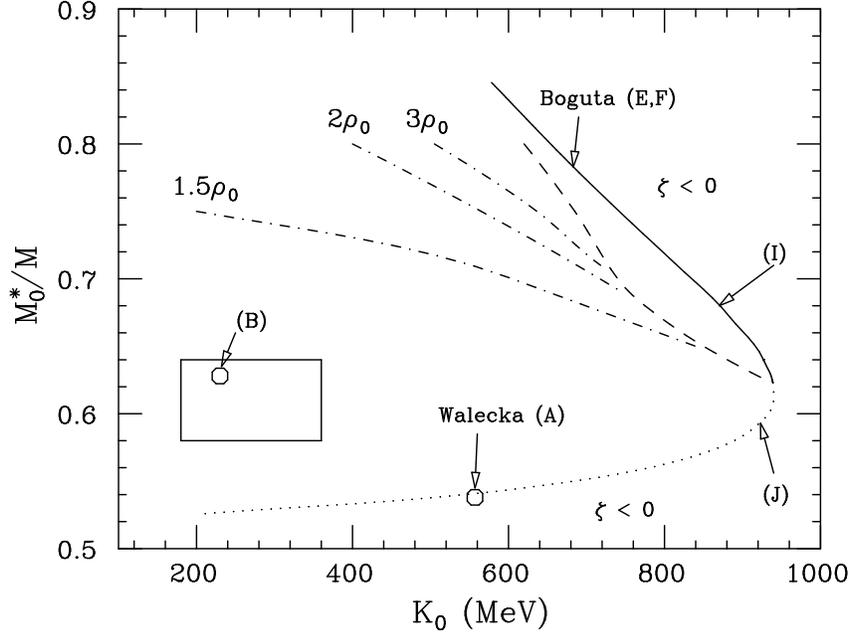}}
\vspace*{.1in}
\caption{\capcrunch{\protect\footnotesize%
 $\Mstarzero/M$ vs.\ $\Kzero$ for
generalizations of the conventional linear sigma model.
All results have $\ezero = -16.1\MeV$ and $\rhozero =
0.1484\fm^{-3}$.
The box denotes the desirable region, and the results of some
specific parametrizations are indicated by capital letters.
The curves are discussed in the text.%
}}

\label{fig:chiralplane}
\end{figure}


The results are summarized in Fig.~\ref{fig:chiralplane}, where some points
corresponding to specific models%
\footnote{These model
parameters from Ref.~\cite{FURNSTAHL93} have been adjusted slightly so that
nuclear matter saturates at $\ezero=-16.1\MeV$.}
from Tables~\ref{tab:one} and \ref{tab:two} are indicated.
(See Table~\ref{tab:twop} for parameter values.)
The solid and dotted curve is the locus of solutions with $\zeta=0$;
on the solid branch, the normal nuclear matter solutions exist at
all densities, while on the dotted branch, the normal solutions
disappear at $\rho \leq 1.38 \rhozero$.
(There is indeed a discontinuity in the behavior of the solutions.)
Regions of the plane to the right and below this curve require
$\zeta<0$, which is physically unacceptable, as discussed earlier.
The dashed curve marks the boundary where normal solutions with
finite $\zeta$ remain stable at all densities; thus, only in the
small region of the plane between the dashed and solid curves can
such solutions be found, and this region lies far from the desired one.

The dot-dashed curves are labeled by the value of the density at
which the normal solution with the specified inputs disappears.
Therefore, while it is technically possible to find parameter sets
for inputs lying within the box, the normal solution disappears at
much too low a density to be useful.
(For example, chiral-model parameters that reproduce the inputs at
point $B$ generate a normal solution that vanishes at
$\rho=1.07\rhozero$.)

Although the choice of limiting density at which
the normal state should
exist is somewhat subjective, it is reasonable to demand that it persist at
least up to $\rho = 2 \rhozero$, since such densities have clearly been
created in the laboratory.
As indicated in the figure, enforcing such a constraint makes it impossible
to achieve the desired values of $\Kzero$ and $\Mstarzero$.
The underlying problem is the $\Phi^3$ coefficient, with its fixed
sign and large magnitude [see Eq.~(\ref{eq:kappabarch})], which yields too
much attraction.
{}From Fig.~\ref{fig:one}, we see that $\kappabar<0$ solutions
(with $\alphabar=\zeta=0$) correspond to values of
$\Kzero$ and $\Mstarzero$ far from the phenomenologically
desirable values.
Allowing $\alphabar=\alphabarp>0$ and $\zeta>0$ can produce
normal solutions with the desired equilibrium properties, but
the consequent nonlinearities are too extreme, and the normal solution
disappears at too low a density.

\mediumtext
\begin{table}[t]
\caption{Parameters for the
 models from Ref.~\protect\cite{FURNSTAHL93} considered in the figures,
 given in the form of dimensionless ratios as in
  Eq.~(\protect\ref{eq:scaledstuff}).  All parameter
  values have been multiplied
  by $10^{3}$ and $\Kzero$ is in MeV.
The coefficients have been adjusted slightly so that $\ezero=-16.1\MeV$.
 All models  have $\zeta=0$, and $\alphabar=\alphabarp$.}
\smallskip
\begin{tabular}{cddddddd}
 Model & \multicolumn{1}{r}{$\textstyle 1 \over \textstyle 2\cs^2 M^2$} &
   \multicolumn{1}{r}{$\textstyle 1 \over \textstyle 2\cv^2 M^2\mathstrut$} &
   $\textstyle \kappabar \over \textstyle 6 M$ &
   $\textstyle \lambdabar \over \textstyle 24$ &
   $\alphabar$ & $\Mstarzero/M$ & $\Kzero$ \\[4pt] \hline
 A & 1.389 & 1.815 & 0.0 & 0.0 & 0.0 & 0.54 & 560 \\[4pt]
 B & 1.478 & 2.309 & 0.940 & $-$0.900 & 0.0 & 0.63 & 230 \\[4pt]
 E,F & 3.499 & 7.315 & $-$3.499 & 0.875 & 14.627 & 0.78 & 680 \\[4pt]
 I & 2.781 & 4.654 & $-$2.781 & 0.695 & 7.061 & 0.68 & 870 \\[4pt]
 J & 2.470 & 3.840 & $-$2.480 & 0.618 & 5.406 & 0.59 & 920 \\[4pt]
\end{tabular}
\label{tab:twop}
\end{table}


\begin{figure}
\setlength{\epsfxsize}{4.5in}
\centerline{\epsffile{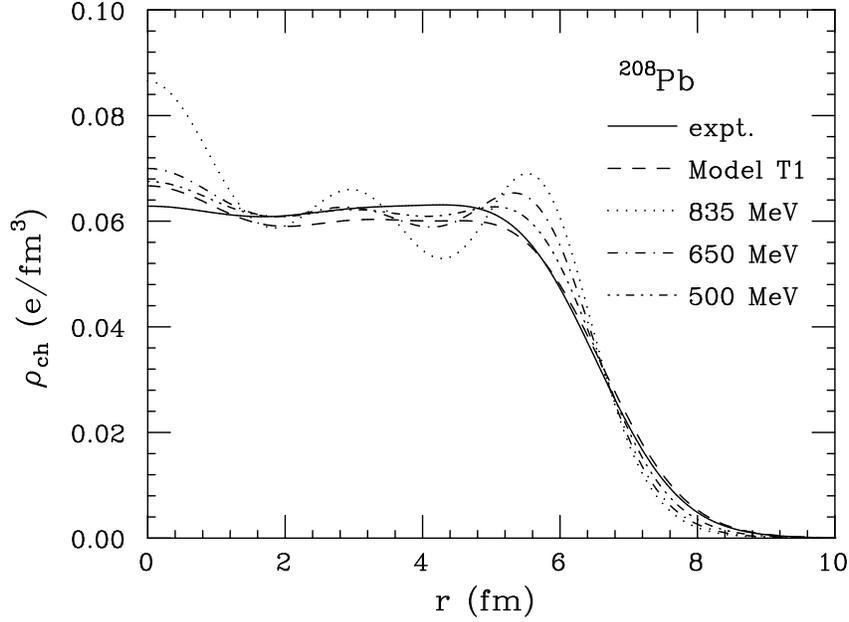}}
\vspace*{.1in}
\caption{\capcrunch{\protect\footnotesize%
Charge densities for chiral model E or F from
Table~\protect\ref{tab:twop} with three different scalar masses
are compared to the density extracted
from experiment.
For comparison,
model~T1 from Ref.~\protect\cite{FURNSTAHL95}, which provides a good
fit to properties of finite nuclei, is also shown.%
}}

\label{fig:eight}
\end{figure}



\begin{figure}
\setlength{\epsfxsize}{4in}
\centerline{\epsffile{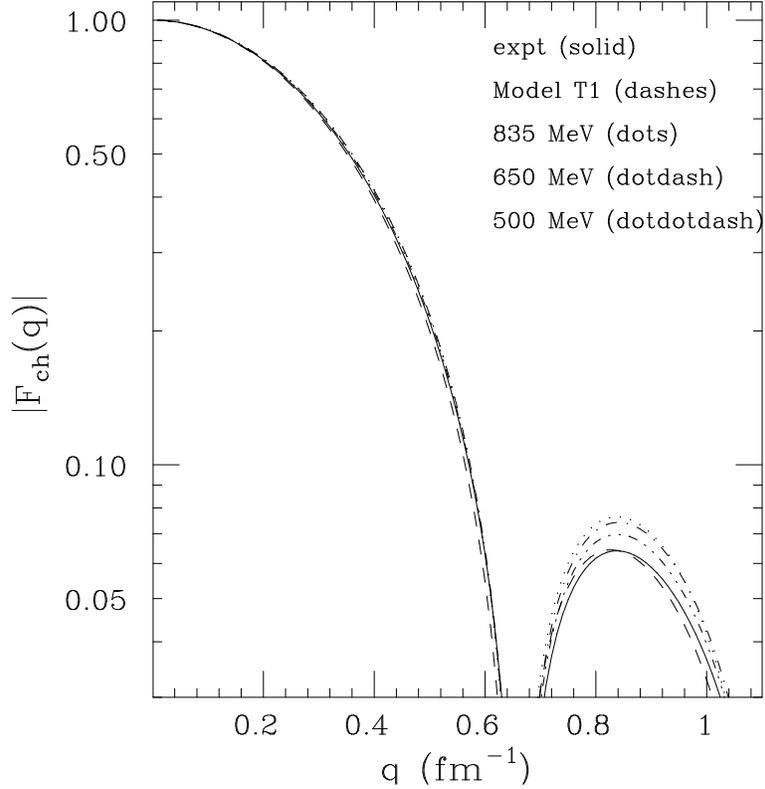}}
\vspace*{.1in}
\caption{\capcrunch{\protect\footnotesize%
Charge form factors $|F_{\rm ch}(q)|$ for chiral model E or F from
Table~\protect\ref{tab:twop} with three different scalar masses
are compared to model~T1 from
Ref.~\protect\cite{FURNSTAHL95} and experiment.
The form factors are simply computed as the Fourier transforms of the
charge densities in Fig.~\protect\ref{fig:eight}.%
}}

\label{fig:eightp}
\end{figure}


For completeness, we also consider the effect of the one-baryon-loop
zero-point energy.
This is strictly applicable only with a renormalizable subset of the
lagrangians (namely, those with $\alphabar=\zeta=0$), but it has been
applied more widely in actual calculations \cite{SARKAR85,FURNSTAHL93}.
The change to the energy density is the addition of
\begin{equation}
  \Delta{\cal E}_c = {M \over 3\pi^2} \Phi^3
       - {1\over 3\pi^2} \Phi^4 + \Deltaevac(\Mstar)  \ ,
     \label{eq:deltaE}
\end{equation}
where $\Deltaevac(\Mstar)$ is the one-loop energy defined in
Ref.~\cite{SEROT86}, which starts at $O(\Phi^5)$ and is numerically
unimportant here.
Historically, the motivation for adding $\Delta{\cal E}_c$ was
to include some repulsion $\propto \Phi^3$ to cancel the strong attraction
in the linear sigma model.
It is now possible to find a saturating solution, which exists at all
densities (see Fig.~59 in Ref.~\cite{SEROT86}), but it is far from
the favored region, since $\Mstarzero/M \approx 0.9$ is much too large.
Allowing additional parameters does not change this result.

Of course, the real test of any parameter set comes in its predictions
for the properties of finite nuclei.
We solve the equations for the finite system using conventional
methods \cite{HOROWITZ81,FURNSTAHL87}.
For chiral models with a linear realization of the symmetry,
we require that
$\gs=\gpi \approx 13.4$ or%
\footnote{We allow the second possibility to compensate for the tree-level
value of $\gA=1$ within the model.}
$\gs=\gpi/\gA$ with $\gA \approx 1.26$.
Thus, once we establish $\cs$ from the nuclear matter analysis,
$\ms$ is determined  (recall that $\cs=\gs/\ms$).
As noted in Ref.~\cite{FURNSTAHL93}, this has dire consequences,
since $\ms$ is always found to be over $600\MeV$,
while phenomenologically successful models typically have
$\ms \approx 500$ to $550\MeV$.
Figures~\ref{fig:eight} and \ref{fig:eightp} show the effects of varying
the scalar mass from $500\MeV$ to $835\MeV$ with the nuclear matter
properties held fixed (models E and F);
we relax the preceding constraint on $\gs$ to obtain
the desired $\ms$.
The vector mass $\mv$ is also held fixed at its experimental value.
Large scalar masses produce oscillations in the charge density
(Fig.~\ref{fig:eight}) or
enhancements in the charge form factor (Fig.~\ref{fig:eightp}),
which are not observed experimentally.
Note that the experimental error bars in Fig.~\ref{fig:eightp} are too
small to include in the figure;
the chiral curves are many standard deviations from the data at
the second maximum.
Moreover, the most unfavorable result shown ($\ms=835\MeV$) is the only
one with an acceptable coupling, namely, $\gs\approx\gpi/\gA$; using the
tree-level value $\gs=\gpi$ produces $\ms=1055\MeV$.

\begin{figure}[p]
\setlength{\epsfxsize}{4.5in}
\centerline{\epsffile{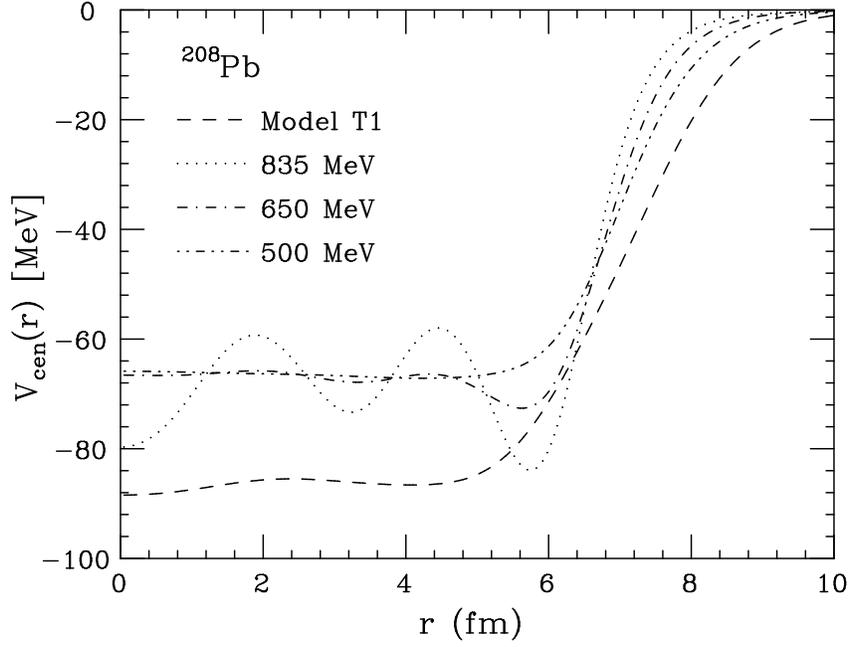}}
\vspace*{.1in}
\caption{\capcrunch{\protect\footnotesize%
 Schr\"odinger-equivalent central potentials
in $^{208}\mbox{Pb}$ for model T1 from Ref.~\protect\cite{FURNSTAHL95}
and a chiral model (E,F) with three scalar masses $\ms = 835\MeV$, $650\MeV$,
and $500\MeV$.%
}}

\label{fig:central}
\end{figure}

\begin{figure}[p]
\setlength{\epsfxsize}{4.5in}
\centerline{\epsffile{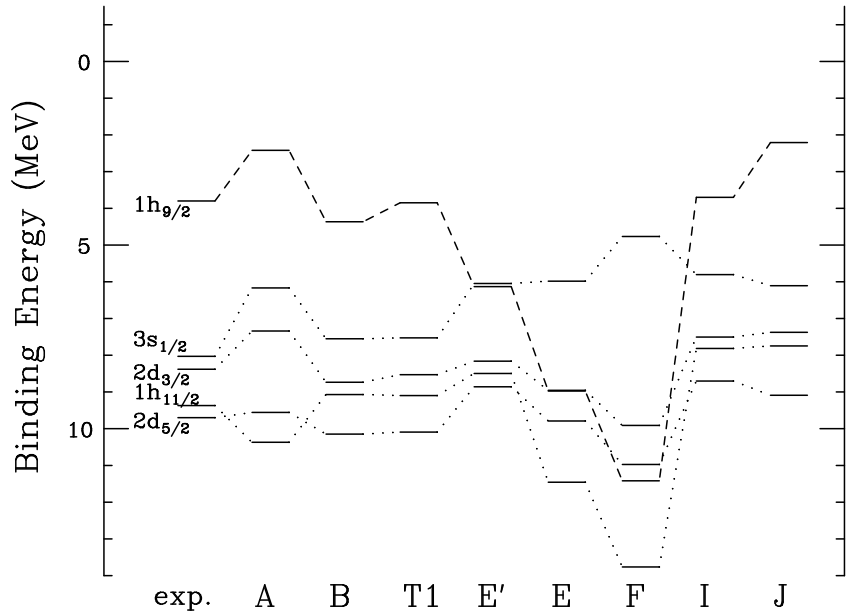}}
\vspace*{.1in}
\caption{\capcrunch{\protect\footnotesize%
 Proton energy levels near the
Fermi surface in $^{208}\mbox{Pb}$.
Model T1 is from Ref.~\protect\cite{FURNSTAHL95} and parameters for
the others are given in Table~\protect\ref{tab:twop} and in the text.
(The scalar mass $\ms = 500\MeV$ in models I and J.)%
}}

\label{fig:levels}
\end{figure}

Other deficiencies are clearly revealed in calculations of finite nuclei,
and this serves to validate our choice of desirable nuclear matter
properties.
Consider, for example, the single-particle spectrum near the Fermi surface
in heavy nuclei.
The large scalar masses required by the known value of $\gs$ produce
significant changes in the effective central potential seen by the
nucleons.
This is illustrated in Fig.~\ref{fig:central}.
As the scalar mass increases, the slope of the potential in the nuclear
surface becomes steeper; indeed, if one ``averages out'' the central
oscillations, the dotted potential is essentially a square well.
Thus, as the scalar mass increases, levels with large (nominal) values of the
orbital angular momentum $\ell$ become more deeply bound.
This is illustrated in Fig.~\ref{fig:levels} by models E$'$,
E and F, which have
scalar masses of $500\MeV$, $650\MeV$, and $835\MeV$, respectively;
the $1h_{9/2}$
level, which should be unoccupied, is pushed down into the highest filled
shell.
Moreover, the large value of $\Mstarzero / M \approx 0.8$ in these models
reduces the spin-orbit splittings considerably, as can be seen by comparing
the $1h_{9/2}$--$1h_{11/2}$ splitting in models E and F with the spectra in
the successful models B and T1 (from Ref.~\cite{FURNSTAHL95}).
The conclusion is that the large values of $\ms$ and $\Mstarzero /M$ obtained
in the chiral models preclude a correct description of the shell closures
in heavy nuclei.


If we are desperate to obtain a reasonable model of finite nuclei,
we might abandon the connection to pion physics by decoupling
$\gs$ from $\gpi$, so that $\ms$ can be set near $500\MeV$.
This will produce reasonable charge densities and effective central
potentials, as verified by Figs.~\ref{fig:eight} and \ref{fig:central}.
Nevertheless, the models are still restricted to the limited accessible
region in Fig.~\ref{fig:chiralplane}, and the consequences are severe.
The best that can be done is to reduce $\Mstarzero/M$ as much as possible,
so that the spin-orbit splittings are improved, and the single-particle
level structure is reasonable.
This is illustrated in Fig.~\ref{fig:levels} by model I (which exists
at all densities) and model J (which disappears at $\rho \approx
5\rhozero /4$); the lower values of $\ms$ and $\Mstarzero$ produce an
acceptable level ordering and shell closure.

\begin{figure}
\setlength{\epsfxsize}{4.5in}
\centerline{\epsffile{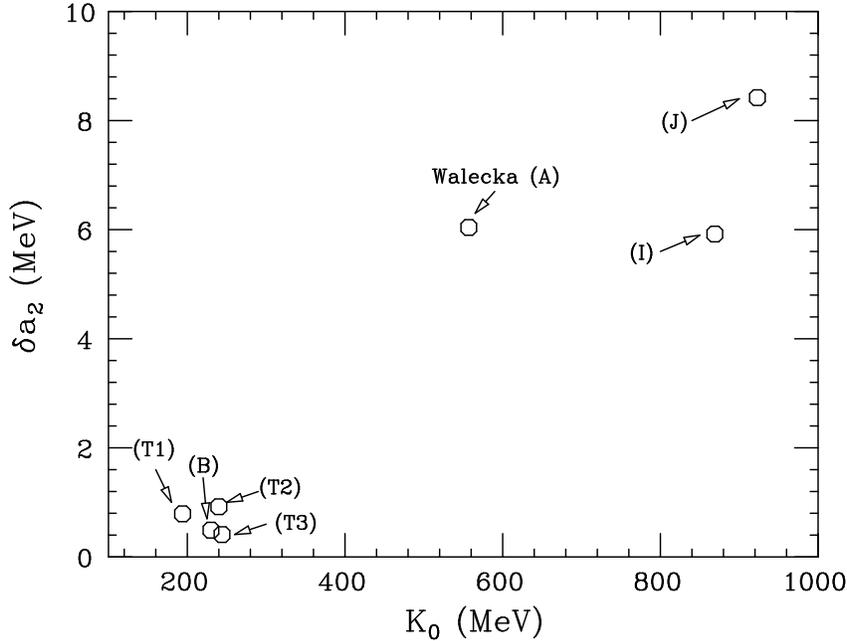}}
\vspace*{.1in}
\caption{\capcrunch{\protect\footnotesize%
 Deviations in the surface energy
$\delta a_2$ (see text) {\it vs.\/} the compression modulus for
the models T1, T2, and T3 from Ref.~\protect\cite{FURNSTAHL95}
and selected models from Table~\protect\ref{tab:twop}.%
}}

\label{fig:six}
\end{figure}

Unfortunately, the tradeoff in these models is also clear: the compression
modulus $\Kzero$ gets very large, and this disrupts the binding-energy
systematics.
This is illustrated in Fig.~\ref{fig:six}, where the deviation in the
calculated surface energy ($\delta a_2$) is plotted as a function of the
compression modulus for different models.
Here $\delta a_2$ is determined by fitting the difference in the
calculated%
\footnote{A center-of-mass correction is applied in determining
the calculated binding energies \protect\cite{REINHARD89}.}
and experimental binding energies of ${}^{16}$O, ${}^{40}$Ca, and
${}^{208}$Pb to a form $\delta E = (\delta a_1) B + (\delta a_2) B^{2/3}$.
(We find $\delta a_1 < 1\MeV$ for all models.  Other analyses give
qualitatively similar results.)
Apparently, models with compression moduli in the desirable range yield
accurate surface energies, while the chiral models I and J clearly yield
surface energies that are too large, roughly 40\% larger than the observed
value ($a_2 \approx 18\MeV$).


Thus, by mapping out the accessible region in parameter space and by
characterizing the dependence of the nuclear observables on $\Mstarzero$,
$\Kzero$, and $\ms$, we conclude that models built upon the conventional
linear sigma model and which feature a ``Mexican-hat'' potential
cannot reproduce
basic nuclear ground-state phenomenology at the Hartree level.
The  nuclear matter analysis of the
simplest version (with $\alphabar=\alphabar'=\zeta=0$) shows that the
constraints on the scalar self-couplings forced by the form of the scalar
potential are
incompatible with desirable equilibrium properties.
Attempts to generate better properties by allowing
$\alphabar=\alphabar'>0$
and $\zeta>0$ inevitably lead to
Lee--Wick solutions with lower energy and to the absence of a normal
solution at densities not far above equilibrium.
The large value of $\gs$ forced by the chiral constraint is an additional
problem, since it requires the scalar mass to be too large.

A different realization of the chiral symmetry seems more compatible with
observed nuclear properties and with successful relativistic mean-field
models.
Here one takes the chiral scalar mass to be {\em large\/} to eliminate the
unphysical scalar nonlinearities and then generates the mid-range attractive
force between nucleons {\em dynamically\/} through correlated two-pion
exchange \cite{JACKSON75,DURSO77,DURSO80,LIN89,LIN90}.
In principle, this approach can be realized within the models discussed
above, albeit at the expense of much greater complexity, since one must
first construct a boson-exchange kernel containing correlated two-pion
exchange and then allow this kernel to act to all orders (for example, in a
ladder approximation) to determine the NN interaction and the resulting
nuclear matter energy density \cite{MACHLEIDT89}.
This calculation is further complicated by the necessity of maintaining
chiral symmetry at finite density, which is difficult to do when one uses
the non-derivative (``pseudoscalar'') $\pi$N coupling implied by the linear
realization of the symmetry \cite{MATSUI82,HOROWITZ82,WALECKA95}.

Alternatively, this approach can be implemented more easily with a
{\em nonlinear\/} realization of the chiral symmetry.
If desired, the mass of the heavy chiral scalar can be kept finite, so
that it plays the role of a regulator that maintains the renormalizability of
the model \cite{MATSUI82}, or the mass can be taken to infinity, so that
the chiral scalar field decouples, resulting in the effective nonlinear
model of Weinberg \cite{WEINBERG67}.
The strong scalar-isoscalar two-pion exchange can be simulated by adding a
low-mass, ``effective'' scalar field coupled directly to the
nucleon, and scalar self-interactions can be added to include a density
dependence in the mid-range NN attractive force.
The chiral symmetry is left intact, because the light scalar is an
isoscalar, and the transformation rules on the nucleon field are
nonlinear \cite{WALECKA95}.
Since the pion mean field vanishes, the resulting chiral mean-field theory
produces an energy functional just like the ones considered in
Sec.~\ref{sec:setup}.
Moreover, although the coupling strength $\gs$ of the light scalar is
comparable to $\gpi$ (as verified by explicit calculation of the correlated
two-pion exchange \cite{LIN89,LIN90}), we no longer require
$\gs \! = \! \gpi$ and are free to adjust $\gs$ within a reasonable range.
In short, the general energy functionals studied earlier (including the
Walecka model) are consistent with the underlying chiral dynamics of QCD,
although its realization in nuclear physics is subtle.
In the next section, we illustrate this scenario with a specific example
from Ref.~\cite{FURNSTAHL95}.


\section{Chiral Models that Work}
\label{sec:tang}

Here we discuss a chiral model that {\it can\/} successfully describe finite
nuclei at the Hartree level.
As we have seen, variants of the linear sigma model, with spontaneous
symmetry breaking from a ``Mexican-hat'' potential, impose strong dynamical
assumptions on the interactions of the light scalar.%
\footnote{A linear sigma model variation, with spontaneous symmetry breaking
from a logarithmic potential, {\it can\/} provide a good description of finite
nuclei, as shown in Ref.~\cite{HEIDE94}.}
In contrast, by working with a nonlinear representation of chiral symmetry,
we remove these constraints, because the chiral-singlet scalar
(which is called $\sigma'$ by Weinberg \cite{WEINBERG67}) can be
decoupled from the other fields.
We can then introduce a {\em new\/} chiral scalar to incorporate the
dynamics in
the scalar-isoscalar sector, as discussed at the end of the last section.

We consider the lagrangian from Ref.~\cite{FURNSTAHL95}:
\begin{eqnarray}
{\cal L}(x) &=&
         \overline N \Bigl(i\gamma^{\mu} {\cal D}_{\mu}
            + \gA\gamma^{\mu}\gamma_5
             a_{\mu} - M  + \gs\phi + \cdots \Bigr)N
       -\frac14 (\partial_\mu V_\nu - \partial_\nu V_\mu)^2
                              \nonumber \\
 &  &  \null  + \frac12 \biggl[
      1+ \eta {\phi\over \Szero}  + \cdots \biggr]
         \Bigl[ {1\over 2}f_{\pi}^2\, {\rm tr}\,
                  (\partial_{\mu}U\partial^{\mu}U^{\dagger})
                + \mv^2 V_{\mu}V^{\mu} \Bigr]
                                       \nonumber \\
         &  &       \null    +{1\over 4!}\zeta
             (\gv^2 V_{\mu}V^{\mu})^2
             + {1\over 2}\partial_\mu \phi \partial^\mu \phi
         - H_{\rm q}\biggl({S^2 \over \Szero^2}
         \biggr)^{2 / d}    \biggl( {1 \over 2d}
              \ln {S^2 \over \Szero^2}
                   -{1 \over 4} \biggr)      + \cdots
                \ ,\label{eq:NLag}
\end{eqnarray}
where  $\gA \approx 1.26$ is the axial coupling constant,
${\cal D}_{\mu} = \partial_\mu + iv_\mu + i \gv V_\mu$
is a chirally covariant derivative, and $U$, $v_\mu$, and $a_\mu$ depend
on the pion field (see Ref.~\cite{FURNSTAHL95} for details).
Here we write the nucleon field as $N$ to distinguish it from the field
$\psi$ used previously, because $N$ transforms nonlinearly under the
chiral symmetry.
A novel feature of this model is that the scale dimension $d$ of the
new scalar field $S(x)$ is allowed to differ from unity.
The scalar fluctuation field $\phi$ is related to $S$ by
$S(x)\equiv\Szero - \phi(x)$.
The form of the lagrangian is motivated by requiring that the model
satisfy the low-energy theorems of broken scale invariance in QCD at the
tree level in the effective scalar field.
See Ref.~\cite{FURNSTAHL95} for more discussion of the motivation and
consequences of Eq.~(\ref{eq:NLag}).

This lagrangian illustrates one way to specify vacuum contributions,
which at one-baryon-loop order modify all powers of $\phi$.
The polynomial terms in $\phi$
must be combined with corresponding counterterms; in this way,
the vacuum contributions are absorbed into the
renormalization of the scalar polynomial.
In principle, there are an infinite number of unknown parameters.
However, if one insists that the low-energy theorems are satisfied at tree
level in the meson fields,
the end result for the scalar potential must be of the form in
Eq.~(\ref{eq:NLag}), where the couplings are renormalized
\cite{FURNSTAHL95}.%
\footnote{Another way to include vacuum effects
is to insist on the renormalizability of the lagrangian.
Then the first few powers of $\phi$ are fixed by a set
of renormalization conditions, and the higher powers contain finite,
calculable coefficients.}
This potential can be expanded as a polynomial in $\phi$
and {\it all\/} coefficients are determined by the three parameters
$S_0$, $d$, and $H_q = \ms^2 d^2 S_0^2 /4$.
One never has to explicitly calculate any counterterms;
the nucleon-loop effects are automatically included in the scalar potential
contained in Eq.~(\ref{eq:NLag}).
The assumption of naturalness provides an alternative (but not inconsistent)
justification for limiting the
impact of the quantum vacuum at ordinary density
to the renormalization of a few parameters, as discussed below.

Thus we have only a few unknown renormalized constants (parameters);
in Ref.~\cite{FURNSTAHL95} these were determined by fitting directly
to finite nuclei, using a $\chi^2$ minimization algorithm.
Although we will not reproduce the results here, the charge densities,
single-particle spectra, and binding-energy systematics are in good
agreement with experiment.
Moreover, the predicted nuclear matter equilibrium properties ($\ezero$,
$\rhozero$, $\Mstarzero$, and $\Kzero$) fall within the desired ranges.

To connect this model with the present analysis, we present
the energy density $\ed$ for uniform nuclear matter in the Hartree
approximation:
\begin{eqnarray}
     \ed(\phi,\Vzero;\rhoB)
        &=& {1\over 4} m_{\rm s}^2 S_0^2 d^2
           \Bigl\{ \Bigl(1-{\phi\over S_0}\Bigr)^{4/d}
                 \Bigl[{1 \over d}\ln \Bigl(1-{\phi\over S_0}\Bigr)
                    -{1\over 4}\Bigr]
                  + {1\over 4} \Bigr\}
           +g_{\rm v}\rho_{\scriptscriptstyle\rm B}
            V_0 - {1\over 4!}\zeta
             (g_{\rm v} V_0)^4
                   \nonumber \\[4pt]
     & &         - {1 \over 2}\Bigl(1+\eta {\phi_0\over S_0}\Bigr)
                   \, \mv^2
             V_0^2
     + \edk(\phi;\rhoB)
           \ ,         \label{eq:endens}
\end{eqnarray}
The parameter sets T1, T2, and T3
determined in Ref.~\cite{FURNSTAHL95} exhibit naturalness (see Sect.~VI),
as defined earlier; thus, if we expand the scalar potential in powers of
$\phi$, terms beyond $O(\phi^4)$ are  unimportant (although not negligible)
at nuclear
densities and below, where $\Phi/M \lesssim 0.4$.
With this truncation, the energy density becomes a special case of
the general form considered earlier, and
expressions for $\kappabar$,  $\lambdabar$, and $\alphabar$ are given
in Table~\ref{tab:two}.

This example illustrates that the general
functional in Eq.~(\ref{eq:efunctional}) is consistent with the
spontaneous breaking of chiral symmetry in QCD in a nonlinear realization.
The key point is that chiral symmetry imposes no constraints on the
scalar potential or the scalar--nucleon interaction, if a chiral-singlet
scalar field generates the midrange attraction.
(Of course, all interaction terms must be isoscalars, and the
nucleons must obey a nonlinear chiral transformation \cite{WEINBERG67}.)
The lagrangian in Eq.~(\ref{eq:NLag}) provides an explicit example of
a model that is consistent with chiral symmetry (and broken scale
invariance) and that produces accurate results for finite nuclei in
the Hartree approximation; moreover, the coefficients are natural,
thus allowing
for a truncation of the energy functional.
In contrast to linear chiral models with ``Mexican-hat'' potentials,
the logarithmic potential in Eq.~(\ref{eq:endens}) allows for relatively
weak scalar nonlinearities and the dominance of the cubic and quartic terms
with reasonable values of the scaling dimension $d$;
the coefficients of the scalar terms are natural as a result.
Although the logarithmic form was motivated by considering the broken scale
invariance of QCD, we could equally well impose as a {\em phenomenological
principle\/} that the coefficients in the energy functional be natural,
since the subsequent truncation leads to similar results.
In short, the energy functionals considered in Sec.~\ref{sec:setup}
are consistent with spontaneously broken chiral symmetry, the broken
scale invariance of QCD, and the basic observed properties of finite
nuclei.


\section{Discussion}
\label{sec:discuss}

Descriptions of nuclear matter and finite nuclei, which are ultimately
governed by the physics of low-energy QCD,
are efficiently formulated using low-energy degrees of freedom,
namely, the hadrons.
The application of effective field theory methods to nuclear physics
problems, however, is in its adolescence, with new developments
occurring rapidly.
In this work, we try to build upon the success of relativistic
mean-field phenomenology by using ideas from density functional
theory and effective field theory.
One of the key issues is {\em naturalness}.

As noted earlier, we have excluded many terms from our energy
functional: higher-order
polynomials in the vector fields and mixed scalar--vector terms,
derivative terms, and so on.
In retrospect, were we justified in neglecting them?
An important assumption in applying effective field theories,
such as chiral perturbation theory, is that the coefficients of terms in the
lagrangian are ``natural'', {\em i.e.}, of order unity,
when written in appropriate dimensionless units.
This assumption permits the organization of terms through a
power-counting scheme, because one can systematically
truncate the expansion when working to a desired level of
accuracy \cite{GEORGI93}.
We  propose an analogous concept of naturalness for the energy functional,
which will justify the neglect of higher derivatives and
powers of the fields when applying Eq.~(\ref{eq:efunctional}) to nuclei.
This is the basic idea that underlies our effective field theory approach
to the nuclear many-body problem.

In performing a mean-field analysis \cite{BODMER91,FURNSTAHL96}, one
can easily identify dimensionless ratios that set the scale of individual
contributions to the energy.
For example,
one can rewrite the scaled energy density of nuclear matter,
$\ed/M^4$, in terms of the dimensionless ratios
$g_{\rm v}V_0 /M = W/M$ and $g_{\rm s}\phi /M = \Phi/M$,
which then become our finite-density expansion parameters.
If one expresses the nuclear matter energy density in terms of
the scaled field variables written above [see
Eqs.~(\ref{eq:bds15})--(\ref{eq:bds17})], one finds that the ratios
\begin{equation}
    {1 \over 2 \cv^2 M^2} \ , \quad
    {1 \over 2 \cs^2 M^2} \ , \quad
    {\kappabar \over 6 M} \ , \quad
    {\lambdabar \over 24} \ , \quad
    {\zeta \over 24}\ , \quad
    \alphabar \ , \quad
    \cdots
       \label{eq:scaledstuff}
\end{equation}
should all be of roughly equal size (or at least of the same order of
magnitude) for our expansion to be ``natural''.

In Table~\ref{tab:three}, we illustrate these ideas using parameter
sets T1, T2, and T3 from Ref.~\cite{FURNSTAHL95}.
These sets accurately reproduce nuclear properties, and we observe that
the parameters are consistent with our definition of naturalness.
The nonlinear scalar parameters tend to be somewhat small, but it
is clear that one can shift strength between terms with equal powers
of the meson fields, which follows because both $\Phi$ and $W$ are roughly
proportional to $\rhoB$.
Thus, to make concrete conclusions about the importance of nonlinearities,
one should include all terms consistent with the level of truncation.
(Note, however, that the quartic parameter $\alphabarp$ was set to zero
in Ref.~\cite{FURNSTAHL95}; when it is included in a fit, the
optimal parameter set becomes {\it more\/} natural.)
Observe also the important result that the highest-order terms retained
($\zeta$, $\lambdabar$) do not dominate the energy.

\mediumtext
\begin{table}[tb]
\caption{Dimensionless ratios of Eq.~(\protect\ref{eq:scaledstuff})
     evaluated in the models of Ref.~\protect\cite{FURNSTAHL95}.
     The absolute values are shown, and all have been multiplied
     by $10^{3}$.}
\smallskip
\begin{tabular}{lcccccc}
 Model & $\textstyle 1 \over \textstyle 2\cs^2 M^2$ &
   $\textstyle 1 \over \textstyle 2\cv^2 M^2\mathstrut$ &
   $\textstyle \kappabar \over \textstyle 6 M$ &
   $\textstyle \lambdabar \over \textstyle 24$ &
   $\textstyle \zeta \over \textstyle 24$ &
   $\alphabar$ \\[4pt] \hline
  T1 & 1.48 & 2.25 & 0.02 & 0.03 & 1.68  & 1.16 \\
  T2 & 1.65 & 2.52 & 0.35 & 0.12 & 1.43  & 1.77 \\
  T3 & 1.34 & 1.95 & 0.32 & 0.13 & 1.44  & 0.31 \\
\end{tabular}
\label{tab:three}
\end{table}

The usefulness of naturalness extends to gradient terms, which
do not contribute in uniform nuclear matter.
Experience with mean-field models applied to finite nuclei
shows that the derivatives
of the fields are small; one finds that typically,
$|\nabla\Phi|/M^2$ and $|\nabla W|/M^2$ are  $(0.2)^2$ or less.
If we assume mean-field dominance, such that fluctuations around
the mean fields are small, and the naturalness of the
coefficients in the derivative expansion,
we can truncate the derivative terms at some tractable order.
In the models studied here
we consider only quadratic terms in gradients of the fields.

Naturalness implies that we should not expect qualitative changes in our
description from additional higher-order terms.
If the physics is consistently dictated by the last terms added,
the system (or framework) is not natural.
Further support for the naturalness assumption in our
framework comes from extending the
model in Ref.~\cite{FURNSTAHL95}
to include $\phi^2 V_\mu V^\mu$ and $(V_\mu V^\mu)^3$ terms and then
repeating an optimization directly to selected properties of finite nuclei.
The new fits are very close to the fits obtained without these terms.
Furthermore, contributions to the energy from the new terms are
less than 10\% of those from the old terms at nuclear
matter density, and the old coefficients change only slightly in the
new fit \cite{FURNSTAHL96}.
Thus contributions from the higher-order terms can be absorbed into
slight adjustments of the coefficients in Eq.~(\ref{eq:NLag}).
Moreover, since the terms quadratic in gradients of the
field are small, one expects that adding
additional field gradients will produce only minor effects.

Naturalness also implies that we can exploit the freedom to make nonlinear
field redefinitions.
Although we have no general proof that finite-density observables are
unchanged under these redefinitions, it is certainly true at the
mean-field level ({\em i.e.}, with classical meson fields).
This freedom allows us to show that our couplings to the nucleon field
are more general than is apparent at first.
For example, if instead of $\gs\psibar \phi\psi$ we have $\gs\psibar
f(\phi)\psi$, with $f(\phi) = \phi + \dots$, we can redefine the scalar
field to recover our standard form.
Similarly, we can reproduce contact interactions among the nucleons by
appropriately choosing nonlinear interactions among the meson fields;
eliminating the meson fields using the field equations then provides the
desired products of nucleon densities and currents.
Thus there are numerous ways to write the energy functional that produce
equivalent results, provided one keeps all orders in the interaction terms.
The basic question is which representation leads to the most
efficient truncation scheme, namely, one with meaningful dimensionless
expansion parameters, that can be truncated at low order and still provide
an accurate description of empirical data, using ``natural'' coefficients.
Whereas we cannot say that other ways of writing the energy functional
will not be useful \cite{SKYRME},
we have at least found one way of parametrizing the
interactions (simple Yukawa meson--nucleon couplings with additional
nonlinear couplings between meson fields) that satisfies the desired
constraints.
These issues will be discussed further in Ref.~\cite{FURNSTAHL96}.

One might instead decide to parametrize the nuclear matter energy density by
simply expanding in powers of the Fermi momentum $\kfermi$ and by
adjusting the coefficients in this expansion to the desired nuclear
matter properties.
In practice, this is neither efficient nor illuminating.
One finds that to reproduce accurately the density dependence found in
successful
models, one must include many terms in the expansion [roughly through
$O(\kfermi^{14})$ in $\ed /\rhoB$].
Moreover, even with this many terms,
extrapolations to higher density are problematic.
If all of the coefficients were independent, they would not be
sufficiently
constrained by the few conditions on nuclear matter near equilibrium.
In contrast, in models with meson fields,
the parameters are correlated, and these constraints are effective in
restricting the relevant region of parameter space.

If one accepts the validity of our assumption of naturalness, we can compare
our truncated energy functional to one obtained directly from the Hartree
(or one-baryon-loop) approximation to a given lagrangian.
Although both approaches lead to Dirac equations for nucleons moving in
local meson fields, the parameters in the energy functional
and in the lagrangian may actually be related in a complicated fashion.
The basic premise of density functional theory is that it is possible to
parametrize exchange and two-nucleon
correlation effects through local (Kohn--Sham)
potentials (or self-energies), and that is how we may interpret our
local meson fields.

This approach is useful, however, only for ground-state properties like
the total energy and particle densities, and in general, the single-particle
wave functions and eigenvalues are simply mathematical constructs that have
no direct relation to observables.
Fortunately, relativistic calculations including exchange and
two-nucleon correlations
show that the Hartree contributions dominate the self-energies, which
are essentially state independent.
Thus we expect that the large scalar and vector fields seen at the Hartree
level will control the dynamics, and that the majority of exchange and
correlation effects can be included by adjusting the nonlinear parameters
(and thereby the density dependence) contained in the energy functional;
this ``Hartree dominance'' also makes it reasonable to compare the
single-particle energies with observed spectra, at least for states near
the Fermi surface.
These considerations again indicate the efficiency of writing the energy
functional in terms of self-interacting meson fields: at normal nuclear
densities these fields are small enough compared to the nucleon mass to
provide useful expansion parameters, yet large enough that exchange and
correlation corrections can be included, at least approximately, as minor
perturbations.

Although this mean-field approach allows for an accurate description of
many nuclear properties and has been applied to various regions of the
Periodic Table, it is really just the first step in the development of a
systematic treatment of the relativistic nuclear many-body problem based
on effective field theory.
There are many limitations and unsolved problems.
First, there is the question of a more complete inclusion of higher-order
corrections in the energy functional.
These could be computed by starting with a lagrangian whose parameters
are fit to NN data and then by calculating at a similar level (for example,
DBHF) at finite density.
Although it is known that one can parametrize the density dependence of
the DBHF self-energies with meson self-interactions \cite{GMUCA92}, and that
one can include rearrangement effects consistently in the single-particle
Dirac equations \cite{LENSKE95}, the state dependence of the self-energies
is neglected in these calculations.
By comparing results of more sophisticated nuclear matter calculations with
their mean-field parametrizations, one may be able to optimize the
description of state-dependent effects, which
will be important if one wants to study the
nucleon--nucleus optical potential at high energy, for example.

The description of the quantum vacuum should also be improved.
As discussed earlier, it is possible to include these effects at the one-loop
level by incorporating them into the model parameters; if our naturalness
assumption is valid, only a few such parameters are needed in practice.
Higher-order contributions, however, will require counterterms involving
fermion fields and will introduce vacuum contributions that explicitly
involve the valence nucleons (the strong Lamb shift, for example).
Moreover, the present approach neglects contributions from meson loops.

One must also consider the limitations of the present model, particularly
with regard to extrapolation to high density.
At nuclear densities and below, we have seen that only a few terms are
needed in the energy functional to describe nuclei, and that the sensitivity
to higher-order products of fields is minimal.
This precludes a determination of the coefficients of these higher-order
terms from empirical data.
As the density increases, however, $\Phi/M$ and $W/M$ increase accordingly,
and the unknown higher-order terms may become important.
This prevents a controlled extrapolation to high densities, such as those
relevant in the interiors of neutron stars.
It is not clear at present what the limits to reliable extrapolation are.
These difficulties are shared by all existing treatments of the nuclear
many-body problem in terms of hadrons; for example, calculations of
few-nucleon systems based on nonrelativistic
potentials \cite{WIRINGA91,CARLSON91} are sensitive
to three-body (and perhaps four-body) forces, but reliable extrapolation
to high density is likely to require information on additional many-body
forces as well.

\begin{figure}
\setlength{\epsfxsize}{4in}
\centerline{\epsffile{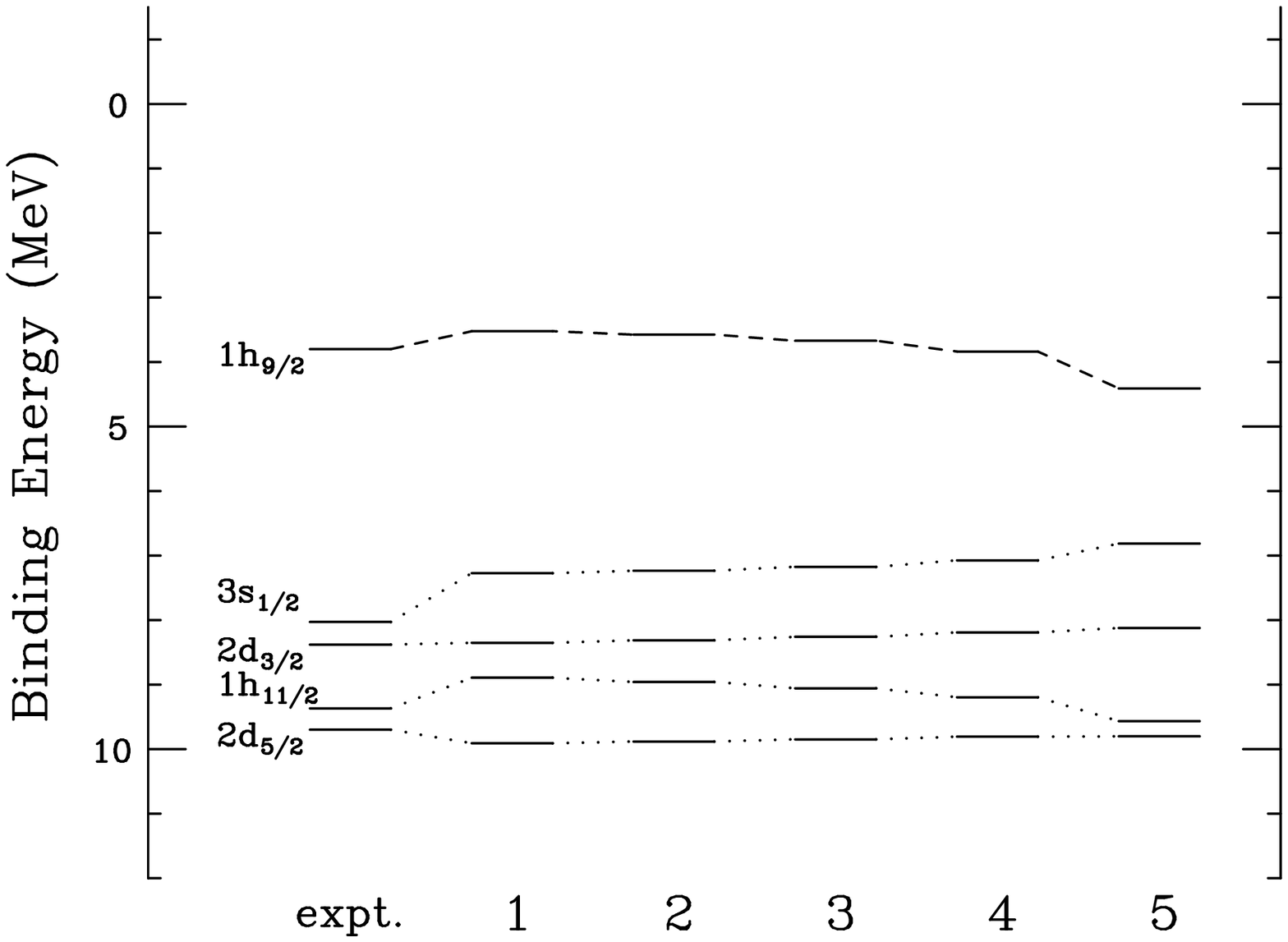}}
\vspace*{.1in}
\caption{\capcrunch{\protect\footnotesize%
 Single-particle proton levels near the Fermi
surface in $^{208}$Pb for five models with fixed
$\ezero$, $\rhozero$, $\Mstarzero$, and $\Kzero$, but
different $(S/K)_0$.%
}}
\label{fig:skewness}
\end{figure}

To close the discussion, we return to the question of whether the nuclear
properties considered here lead to additional constraints on the nuclear
matter saturation curve near equilibrium.
As noted in Sec.~\ref{sec:setup}, some recent work indicates that the
ratio of the ``skewness'' $S$ to the compression modulus $K$ is constrained
by monopole vibrations \cite{PEARSON91,RUDAZ92a}.
To study this issue, we considered an energy functional of the form in
Eq.~(\ref{eq:efunctional}) with $\alpha=\alpha'=0$ and varied the ratio
$(S/K)_0$ by varying $\zeta$
while holding all other nuclear matter inputs fixed.
We found little correlation between the value of $(S/K)_0$ and the nuclear
observables studied here; for example, in Fig.~\ref{fig:skewness},
we show the proton levels near the Fermi surface in ${}^{208}$Pb from
five models for which $(S/K)_0$ ranges by
more than a factor of five.
As long as the parameter set is natural, the effects of varying the skewness
are small, and similar results were found for other observables.
These results support Bodmer's conjecture \cite{BODMER91} that the spin-orbit
splittings in nuclei are determined essentially by $\Mstarzero$ and are
independent of the specific form of the nonlinear interactions.


\section{Summary}
\label{sec:summary}

In this paper, we demonstrate that it is impossible to unite relativistic
mean-field phenomenology with manifest chiral symmetry by building upon
the conventional linear sigma model, which contains a ``Mexican-hat'' potential
and a light scalar meson playing a dual role as the chiral partner of the
pion {\it and\/} the mediator of the intermediate-range NN attraction.
We illustrate the generic failure of this type of model at the Hartree level
as well as the characteristics of chiral models that {\it can\/}
successfully describe finite nuclei.

Although our study of chiral models involves specific lagrangians treated
at the Hartree (or one-baryon-loop) level, the analysis is based on the
more general concept of an energy functional that describes the finite,
many-nucleon system.
The energy functional depends on valence-nucleon Dirac wave functions and
classical scalar and vector meson fields.
We interpret this energy functional within the framework of density
functional theory, which says that the exact energy functional for the
system can be written entirely in terms of the scalar density and baryon
four-current density.
Rather than work solely in terms of densities, we include auxiliary meson
fields, which correspond to (local) Kohn--Sham potentials, and which provide
an efficient parametrization of the density dependence.
Thus, with a sufficiently general energy functional, we can include
many-body effects beyond the simple Hartree level, even though we retain
only classical meson fields and local interactions.
Rather than attempt to compute the form of the functional directly from
an underlying lagrangian, we use observed nuclear properties (and
extrapolations to nuclear matter) to determine the unknown coefficients.

In the general case, the energy functional contains an infinite number of
interaction terms involving the nucleon and meson fields, and their
derivatives.
This is actually an advantage, since transformations of the field variables
allow one to cast the functional in different forms, with the goal being
to find the representation that allows the most efficient truncation.
For the truncation to be plausible, one must have a reasonable (small)
expansion parameter, and one must assume a principle of ``naturalness'',
which says that the coefficients of the expansion, when written in
appropriate dimensionless ratios, must all be of order unity.
These are precisely the basic ideas underlying applications of effective
field theories, such as chiral perturbation theory.
We showed that such a truncation of the energy
functional is possible at and below
normal nuclear densities, if the functional is written in terms of
self-interacting scalar and vector fields with Yukawa couplings to the
nucleons; under these conditions, the ratios of the meson fields (and their
gradients) to the nucleon mass are small enough to serve as useful expansion
parameters, and sets of coefficients can be found that both reproduce
observed nuclear properties and satisfy the naturalness assumption.
In practice, truncation of the energy functional at quartic terms in
the meson fields and at quadratic terms in field gradients is sufficient for
the observables studied here.

To constrain the energy functional, we use nuclear observables that are
meaningful from the standpoint of the underlying density functional
framework, such as (charge) densities and total binding energies.
Although in general, the single-particle Dirac eigenvalues obtained in the
local potentials are not directly related to observables, we note that in
the relativistic nuclear many-body problem, the single-particle potentials
(or self-energies) are known to be dominated by the Hartree contributions
and are essentially state independent.
Thus the exchange and correlation effects included in the potentials are
small, and it is reasonable to also use observed nuclear spectra as input,
at least for states near the Fermi surface.
In short, the observables we use to constrain the parameters are precisely
the ones that should be calculable in a density functional
framework where the Hartree terms dominate the self-energies.
We also observe the fortunate result that the scalar and vector fields are
small enough compared to the nucleon mass to serve as useful expansion
parameters, yet large enough that exchange and correlation effects enter
only as small perturbations (at least for occupied states).

Rather than fit directly to the properties of finite nuclei, we extrapolate
instead and determine our parameters from the properties of nuclear matter
near equilibrium.
In nuclear matter, the energy functional reduces to a function of the Fermi
momentum and the (constant) scalar and vector fields; it is then possible
to invert the field equations and express the unknown parameters directly
in terms of the nuclear matter ``observables''.
This allows us to scan the parameter space to find sets that lead to an
accurate reproduction of nuclear properties.
More importantly, we can exclude models that are too constrained (by
symmetry, for example) by showing that it is impossible for them to
reproduce desirable nuclear matter results.

Parameter sets that accurately reproduce nuclear observables typically
involve small but significant nonlinear meson self-interactions.
These self-interactions introduce additional density dependence beyond that
contained in simple Yukawa meson--nucleon couplings.
This density dependence can be interpreted either in terms of meson masses
that are modified in the nuclear medium or in terms of variations in the
scalar and vector self-energies.
We can make three important observations about this additional density
dependence:
First, it can be parametrized efficiently using self-interactions of the
meson fields.
Second, to make meaningful statements about the size of these
self-interactions, one must include all allowed terms at the given level
of truncation, as required by naturalness.
This is relevant since most relativistic mean-field calculations to date
have concentrated on scalar self-interactions only, omitting scalar--vector
and vector--vector terms.
Third, although we choose to determine these interactions by fitting to
empirical nuclear properties, one could instead fit them to the results
of more detailed nuclear matter calculations based on an underlying
lagrangian, such as DBHF.
If the DBHF self-energies are determined from fits to NN observables, this
alternative procedure provides a direct link between relativistic
finite-nucleus structure calculations and the two-nucleon problem.

To address the chiral models, we considered specific lagrangians
that realize the spontaneously broken chiral symmetry of QCD in different
ways and studied them at the Hartree level.
The resulting energy functionals are special cases of the general one
studied earlier, so we apply our previous analysis and
test whether acceptable parametrizations can be obtained.
We find that the accessible region for linear sigma
models is far from the region of acceptable nuclear phenomenology.
The results of this new analysis solidify our previous
conclusion \cite{FURNSTAHL93} that chiral hadronic models built upon
the conventional linear sigma model cannot reproduce observed properties
of finite nuclei, at least at the Hartree level.
The two critical constraints imposed by the conventional lagrangian are
the ``Mexican-hat'' potential and the connection between the scalar coupling
and the pion coupling ($\gs = \gpi/\gA$).
The former generates scalar self-interactions with the wrong systematics,
and the latter produces a scalar mass that is too large, leading to
unrealistic nuclear charge densities, level orderings, and shell closures.

On the other hand, one can find chiral models that {\it are\/}
phenomenologically successful at the Hartree level.
As an example, we discuss a model that was introduced in a recent paper,
which features a nonlinear realization of chiral symmetry.
One motivation for such a model is that the chiral scalar present in the
linear sigma model is irrelevant for the nuclear dynamics; the mid-range
NN attraction arises instead from the exchange of two correlated pions.
One can simulate this attraction by introducing an auxiliary scalar field
with a Yukawa coupling to the nucleon.
Chiral symmetry is maintained if the nucleon field transforms nonlinearly
and the auxiliary field is a chiral scalar.
Due to the nonlinear realization of the symmetry, the ``Mexican-hat'' potential
is no longer needed, and various constraints on the model parameters
disappear.
At the Hartree level, the energy functional has the full flexibility of the
general functional considered earlier; thus, parameter sets can be found
that accurately reproduce nuclear properties in a framework that is
consistent with the underlying chiral symmetry of QCD.
Moreover, we observe that successful models have parameters that are
consistent with our ``naturalness'' assumption, which forms the basis for
our expansion and truncation scheme.

Although the general mean-field approach studied here leads to excellent
results for the bulk properties of nuclei, there are many questions to be
answered before one has a consistent, systematic treatment of the
relativistic nuclear many-body problem.
A more accurate inclusion of vacuum, exchange, and correlation effects
should be pursued.
The extension of this framework to deal with high-density matter, where the
expansion parameters are no longer small, is also an outstanding problem.
Some of these ideas will be discussed in a future paper \cite{FURNSTAHL96}.


\acknowledgments

We thank P.~Ellis, H.~M\"uller, and J.~Rusnak
for useful comments and stimulating discussions.
We also acknowledge the Institute for Nuclear Theory at the University
of Washington, where part of this
work was carried out.
This work was supported in part by the Department of Energy
under Contract No.\ DE--FG02--87ER40365, the
National Science Foundation
under Grants No.\ PHY--9203145 and PHY--9258270, and the A. P.
Sloan Foundation.


\end{document}